\documentclass{jfm}

\usepackage{graphicx}
\usepackage{natbib}

\ifCUPmtlplainloaded \else
  \checkfont{eurm10}
  \iffontfound
    \IfFileExists{upmath.sty}
      {\typeout{^^JFound AMS Euler Roman fonts on the system,
                   using the 'upmath' package.^^J}%
       \usepackage{upmath}}
      {\typeout{^^JFound AMS Euler Roman fonts on the seystem, but you
                   dont seem to have the}%
       \typeout{'upmath' package installed. JFM.cls can take advantage
                 of these fonts,^^Jif you use 'upmath' package.^^J}%
      }
  \else
  \fi
\fi

\ifCUPmtlplainloaded \else
  \checkfont{msam10}
  \iffontfound
    \IfFileExists{amssymb.sty}
      {\typeout{^^JFound AMS Symbol fonts on the system, using the
                'amssymb' package.^^J}%
       \usepackage{amssymb}%
         
       \let\ge=\geqslant  
      }{}
  \fi
\fi

\ifCUPmtlplainloaded \else
  \IfFileExists{amsbsy.sty}
    {\typeout{^^JFound the 'amsbsy' package on the system, using it.^^J}%
     \usepackage{amsbsy}}
    {}
\fi

\newsavebox{\astrutbox}
\sbox{\astrutbox}{\rule[-5pt]{0pt}{20pt}}

\title[Energetics and thermodynamics of the 
Boussinesq/Anelastic approximations]
{Thermodynamics/dynamics coupling and thermodynamic consistency of 
Boussinesq and anelastic binary fluids with an arbitrary nonlinear equation
of state}

\author[R. Tailleux]%
{R\ls \'{E}\ls M\ls I\ns T\ls A\ls I\ls L\ls L\ls E\ls U\ls X$^1$%
  \thanks{Present address: Department of Meteorology,
University of Reading, Earley Gate, PO Box 243, Reading RG6 6BB, 
United Kingdom. E-mail: R.G.J.Tailleux@reading.ac.uk}}

\affiliation{$^1$Department of Meteorology,
University of Reading, \\ Earley Gate, PO Box 243, Reading, RG6 6BB, UK}

\pubyear{1996}
\volume{538}
\pagerange{119--126}
\date{16 December  2010 and in revised form ??}

\begin{document}

\maketitle

\begin{abstract}

 This paper shows that the energetics of Boussinesq and anelastic
fluids possesses a term that can be identified as the approximation
$\delta W_{ba}$ to the compressible work of expansion/contraction $\delta W
=-P {\rm d}\upsilon$, where $P$ is the pressure and $\upsilon$ is
the specific volume. It follows that Boussinesq and anelastic fluids
admit explicit compressible effects
and conversions between internal energy and mechanical energy,
under the form of apparent changes in gravitational potential
energy resulting from changes in density by diabatic and adiabatic
effects. From the knowledge of $\delta W_{ba}$, the corresponding
approximation to the ``heat'' $\delta Q_{ba}$ can be constructed in
a consistent way by requiring that the Maxwell relationships be
satisfied, ultimately leading to the construction of a well defined
approximation to the internal energy and ultimately of the full
range of known thermodynamic potentials. These properties make it
possible to endow common forms of the
Boussinesq and anelastic approximations with 
fully consistent energetics and thermodynamics, even when diabatic
effects and an arbitrary nonlinear equation of state for a binary
fluid are retained, without loss of accuracy. In that case, it can
be shown that the sum of kinetic energy and enthalpy is a conservative
quantity, which plays the role of the total energy in the Boussinesq
and anelastic approximations for both diabatic and adiabatic motions. 
This implies that gravitational 
potential energy can be regarded as the difference between enthalpy
and internal energy, and hence as a pure thermodynamic property of
the fluid.

\par

An important implication of the present results is to support the
recent suggestion by \cite{Tailleux2009} that the Boussinesq
approximation is capable of describing potentially large conversions
between internal energy and gravitational potential energy in 
turbulent stratified fluids, which
physically seems to require that an active role be played by the
emission and dissipation of acoustic waves by molecular diffusive
heating and cooling, as well as by the associated divergent velocity
field, a surprising result with potentially important implications
for our understanding of turbulent mixing in stratified fluids if
further confirmed. Another implication of the results is to suggest
that the form of the Boussinesq primitive equations currently used
as the basis for a majority of numerical ocean models possesses 
a potentially significant spurious source of momentum, which 
can in principle be corrected by using an alternative and
more physically-based definition of buoyancy in the hydrostatic
approximation.

\end{abstract}

\section{Introduction}
  
  Many fluid flows of interest in engineering and geophysical fluid
dynamics have low Mach number ($M=U_0/c_s$) 
and small relative density variations $(\rho-\rho_0)/\rho_0$, 
where $U_0$ is a typical
velocity scale, $c_s$ is the speed of sound, $\rho$ is the density
and $\rho_0$ a reference density. It has been common practice over
the past century to regard such flows as incompressible or weakly
compressible, where an incompressible fluid is one whose density
dependence upon pressure is eliminated while still possibly 
retaining its dependence upon temperature (and chemical composition
as the case may be), e.g., \cite{Lilly1996}. There has been much effort
in seeking to take advantage of the smallness of the two above
parameters to develop sound-proof reduced sets of equations that
are somehow simpler to study than the fully compressible Navier-Stokes
equations.
Two particular classes of approximations have been particularly influential 
and key to simplifying the numerical and theoretical analysis of low
Mach number fluid flows, and will be under focus in this paper. 
The first one is the Oberbeck-Boussinesq approximation
(after \cite{Oberbeck1879} and \cite{Boussinesq1903}), which in its
most common form retains only the rotational divergence free
component of the velocity field, and treats the density as constant
everywhere except where it multiplies the acceleration of gravity. The 
second one is the Anelastic approximation, e.g., 
\cite{Ogura1962,Lipps1982,Bannon1996,Durran1989,Ingersoll1982,
Ingersoll2005,Pauluis2008}.
Many other sets can be constructed, which are beyond the scope of this
paper, for instance by using low Mach number asymptotics and multi-scale
expansion techniques, e.g., \cite{Muller1998,Klein2009,Klein2010}.
\cite{Davies2003} offers a review of a number of commonly employed 
reduced sound-proof sets of equations, and show how they respectively
represent normal modes on the sphere.

\par

  The main focus of this paper is on how compressible effects and 
the coupling between mechanical energy and internal energy 
(the dynamics/thermodynamics coupling) are represented in the 
Boussinesq/anelastic approximations, which are known to decouple
either fully or almost fully the thermodynamics from the dynamics for
adiabatic motions and a linearised equations of state, 
i.e., \cite{Spiegel1960,Ogura1962}. In the latter case, the 
Boussinesq/anelastic approximations usually admit a well-defined conservative
energy quantity, e.g., \cite{Lilly1996}, but the issue of the energetic and
thermodynamic consistency appears to be much less understood when
diabatic effects and/or a nonlinear equation of state are retained, because 
the thermodynamics/dynamics coupling then becomes less trivial. In this
respect, the oceanographic case is instructive, as oceanographers have
used the Boussinesq hydrostatic approximation in conjunction with a
realistic nonlinear equation of state (including the pressure dependence)
for many decades as the basis for 
numerical ocean general circulation models of the kind used in climate 
studies, without any apparent obvious drawbacks apart from the lack of 
a well-defined and closed energy budget, see \cite{Tailleux2010}.
Since Boussinesq ocean models appear to work well with a ``compressible''
equation of state, one may wonder how essential the
assumptions of ``incompressibility'' or ``weak compressibility'' are in the 
construction of the Boussinesq and anelastic approximations in the first place.

\par

 To clarify this issue, a thermodynamics perspective is useful. Indeed,
from a thermodynamic viewpoint, 
the very idea that it might be possible to simplify a particular equation of state 
to eliminate its dependence upon pressure while retaining its dependence upon
temperature appears to become very dubious if not outright physically meaningless when
diabatic effects are retained. This is because the way changes in density are 
affected by pressure changes depends critically on the particular thermodynamic
transformation undergone by the fluid parcels. 
Take an isothermal transformation for instance. 
 \cite{Tailleux2010} shows that the isothermal compressibility 
 $\gamma = \rho^{-1} \partial \rho/\partial P|_T$ is linked to the 
 adiabatic compressibility $\rho^{-1}\partial \rho/\partial P|_{\eta}
 =1/(\rho c_s^2)$ (his Eq. A.6) by:
\begin{equation}
    \gamma = \left . \frac{1}{\rho}\frac{\partial P}{\partial \rho}
    \right |_{T,S} = \frac{1}{\rho c_s^2} + \frac{\alpha^2 T}{\rho c_p}
   = \frac{1}{\rho c_s^2} + \alpha \Gamma ,
   \label{compressibilities}
\end{equation}
where $\Gamma = \alpha T/(\rho c_p)$ is the adiabatic lapse rate.
The key point here is that setting up the adiabatic compressibility
to zero (by taking the zero Mach number limit $c_s=+\infty$), while
it makes the fluid effectively ``incompressible'' for adiabatic motions,
fails in general to do so for nondiabatic transformations  if the
thermal expansion coefficient $\alpha$ is allowed to remain nonzero.
Note that for seawater, typical values are:
$1/(\rho c_s^2) \approx 4.10^{-10} \,{\rm P_a}^{-1}$, whereas 
$\alpha \Gamma = \alpha^2 T/(\rho c_p) \approx 7.5 \times 10^{-13}
\,{\rm P_a}^{-1}$, using $c_s=1500\,{\rm m.s^{-1}}$, 
$\alpha = 10^{-4}\,{\rm K}^{-1}$, $c_p = 4.10^3\,{\rm J.K^{-1}.
kg^{-1}}$, and $\rho=10^3\,{\rm kg.m^{-3}}$. For these values, 
the limit $c_s=+\infty$ decreases the isothermal compressibility by
about two to three orders of magnitude, so that even if it fails to
fully eliminate compressibility effects for non-adiabatic motions,
it appears nevertheless capable of reducing them considerably.
This being said, if one agrees that the very concept of an
``incompressible'' fluid becomes physically meaningless when
diabatic effects are retained, then one may also agree that
it might not be that essential
to set up the adiabatic compressibility to zero in the 
first place. Historically, the latter approach was originally motivated
as a natural way to filter out sound waves, but it is now recognised
that imposing $c_s=+\infty$ is not necessarily to filter out sound 
waves, as such a filtering can be more simply achieved by using 
a hydrostatically adjusted
pressure in the equation of state for density, as discussed by
\cite{Deszoeke2002}.

\par

  Whether Boussinesq/anelastic models accurately represent the
conversions between mechanical energy and internal energy recently
came to attention in relation
with the question of how strong would the oceanic overturning
circulation and meridional heat transport be, and how much diapycnal 
mixing would then be supported, if it were possible somehow to suppress
the mechanical stirring due to the wind and tides, which
has been a controversial issue for the past decade, as reviewed in
\cite{Tailleux2009}, \cite{Tailleux2010b} and \cite{Hughes2009}.
The resulting configuration is often referred to as horizontal
convection, see \cite{Hughes2008} for a recent review on the subject.
The reason why understanding the nature of the thermodynamics/dynamics
coupling is important in that case is because the steady-state mechanical 
energy balance reduces to:
\begin{equation}
     \int_{V} P \frac{D\upsilon}{Dt}\,{\rm d}m = \int_{V} \rho 
     \varepsilon_K ,
     \label{real_me}
\end{equation}
which shows that in order to estimate the overall viscous dissipation
rate, one has to estimate the overall work of expansion/contraction,
where $P$ is the pressure, $\upsilon$ is the specific volume,
${\rm d}m = \rho {\rm d}V$ is the elementary mass element of a fluid
parcel, and $\rho \varepsilon_K$ is the viscous dissipation rate.
As discussed by \cite{Tailleux2010b}, the issue of estimating $B$
for a fully compressible fluid is a subtle one. As shown by 
\cite{Paparella2002} and others, considerable analytical progress
can be achieved
for a Boussinesq fluid with a linear equation of state, as in that
case, Eq. (\ref{real_me}) becomes:
\begin{equation}
     \int_{V} g z \kappa \nabla^2 \rho \,{\rm d}V = \int_{V} \rho_0
     \varepsilon_K \,{\rm d}V  = \kappa g_0 \left [ \langle \rho 
     \rangle_{bottom} - \langle \rho \rangle_{top} \right ],
     \label{boussinesq_me}
\end{equation}
which can be integrated analytically, where $\kappa$ is the molecular
diffusivity, $g_0$ the acceleration of gravity and the terms within
brackets are the surface area integrated density at the bottom and
top of the oceans respectively. Using typical oceanic values,
\cite{Wang2005} estimated the right-hand side to be $O(15\,{\rm GW})$,
which is at least two orders of magnitude smaller than the mechanical
power input due to the wind and tides for instance. The smallness of
this number have been widely interpreted as evidence that the 
surface buoyancy fluxes could not by themselves be responsible for
an overturning circulation and associated meridional heat transport
of the observed magnitude.

\par

\begin{figure}
\begin{center}
\includegraphics[width=10cm]{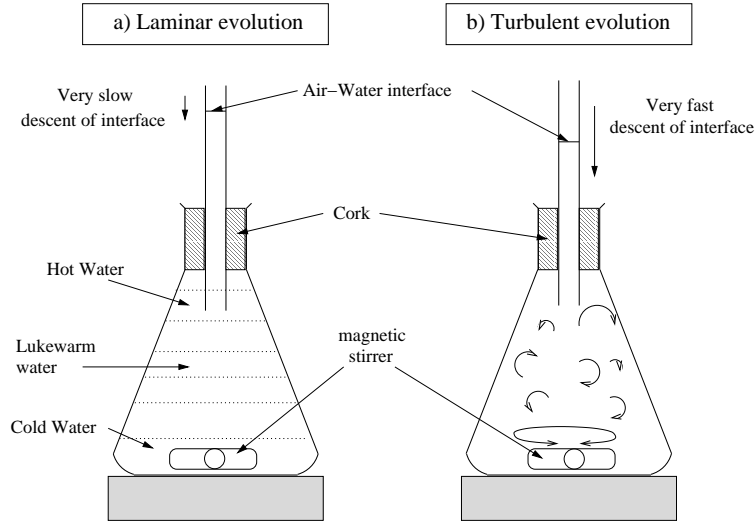}
\caption{Experimental set-up suggesting that compressibility effects 
increase with degree of turbulence in stably stratified fluids.
An Erlenmeyer flask is filled up with water at room temperature.
A vertical stratification is then created by bringing the upper part of
the water near boiling point, and a magnetic stirrer is dropped in the
fluid. The top of the flask is then sealed up 
with a cork through which a thin tube is inserted in order to magnify
the variations of the air-water interface. In laminar conditions 
(panel a), the fluid is near resting conditions, and the air-water
interface moves downward very slowly as the result of molecular diffusion
and/or radiative cooling. Activating the magnetic stirrer generates strong
turbulence (panel b) that results in the spectacular drop of the air-water
interface, presumably as the result of the contraction upon mixing 
discussed in the main text. This experiment was demonstrated to the author
by Peter Rhines during a visit in Seattle in June 2008.}
\label{set_up}
\end{center}
\end{figure}

 The smallness of the r.h.s. of Eq. (\ref{boussinesq_me}) had been previously
interpreted in the context of stratified turbulence
(following \cite{Winters1995}) as implying that conversions between
internal energy and mechanical energy enter the energetics of 
turbulent stratified mixing only at second order. \cite{Tailleux2009} argued,
however, that the apparent smallness of the thermodynamic/coupling in
Boussinesq stratified turbulence
actually hides two large and opposite conversions between internal energy
(IE) and gravitational potential energy (GPE) that almost exactly cancel out, viz.,
one conversion transferring IE into background GPE,
associated with the smoothing out of the vertical reference temperature
gradient and a significant overall volume reduction (assuming that 
$\alpha$ increases with temperature, as is the case for water), the other
conversion dissipating available GPE into IE in a way analogous
to the viscous dissipation of kinetic energy (KE) into IE, and associated with the increase
of the mean thermodynamic equilibrium temperature of the fluid but with
only a negligible overall volume expansion. These results suggest therefore that
compressibility effects must increase with the degree of stratification and
turbulence, which appears to be supported empirically by the kind of
laboratory experiment illustrated in Fig. \ref{set_up}, but have yet to be 
widely accepted, as most subsequent studies so far, e.g., 
\cite{Winters2009,Hughes2009,Nycander2010} continue to favour the
classical view that internal energy and thermodynamics/dynamics coupling
play only a minor role in the energetics of turbulent stratified mixing.

 At least two important misconceptions about weakly compressible fluids 
appear to greatly confuse the understanding of the role of compressible effects
in stratified turbulent fluids.
The first misconception is associated with the widespread (erroneous) belief
that a weak thermodynamics/dynamics coupling is an intrinsic feature of all 
weakly compressible fluids irrespective of the particular thermodynamic 
transformations undergone by fluid parcels, whereas as far as we can judge, 
such a weak coupling is an intrinsic feature
of {\em adiabatic} motions only, with no physical basis for this to be
the case for diabatic motions. The second misconception is associated with the
widespread confusion about what a ``weakly compressible'' or ``incompressible''
fluid is actually supposed to be. Should such a property pertain to a fluid whose
density dependence on pressure is eliminated, as seems to have been the
original intention, e.g., \cite{Lilly1996}, or with the use of the constraints
$\nabla \cdot {\bf v} = 0$ or $\nabla \cdot (\rho_0 {\bf v})=0$, as seems to be
increasingly assumed following the realization that Boussinesq and anelastic
fluids can be used with nonlinear equations of state? Based on the
present analysis, the idea of an incompressible fluid seems to be justified
only for a fluid with zero adiabatic compressibility in the context of purely
adiabatic motions, but physically meaningless otherwise. Physically, the
constraints $\nabla \cdot {\bf v}=0$ and $\nabla \cdot (\rho_0 {\bf v})=0$ do
not imply incompressibility or weak compressibility, in contrast to what is
often assumed, because they do not preclude diabatic or adiabatic density
changes along fluid trajectories. The new view that emerges from the above
arguments is that the Boussinesq and anelastic approximations are actually
able to support potentially large compressible effects and conversions between
internal energy and mechanical energy. If so, this raises many questions that
the present paper seeks to clarify.
1) How do the conversions between internal energy and mechanical energy
in a Boussinesq/anelastic fluid compare with that of a fully compressible fluid?
Does the answer depend on whether diabatic effects and/or a nonlinear 
equation of state are retained?
2) Do the Boussinesq and anelastic approximations conform with the
first and second laws of thermodynamics? What is the form of the
thermodynamic potentials supported by such approximations? How 
different are they from their exact counterparts?
3) Is it an intrinsic problem that many Boussinesq and anelastic models
fail to be energetically and thermodynamically consistent, or can such
models be modified to correct the problem? In the latter case, can this
be done without modifying the formal order of accuracy of the original
Boussinesq/anelastic approximations?

\par

 The main purpose of this paper is to provide an answer to all above
questions, by showing that the Boussinesq and anelastic approximations
can be endowed with well defined energetics and thermodynamics
that closely mimic that of the fully compressible Navier-Stokes equations.
A couple of recent papers have touched upon 
some of these issues. Thus, \cite{Young2010} showed, using ideas
previously developed by \cite{Ingersoll2005}, that the seawater
Boussinesq equation have a well-defined conserved energy quantity
for an arbitrary nonlinear equation of state, although whether this
extends to the diabatic case is not clear. \cite{Pauluis2008} 
addressed a similar issue for the anelastic approximation for a
binary fluid such as moist air, also with a highly nonlinear
equation of state, and discussed energy and thermodynamic consistency
issues. The above studies, however, did not clearly address the
nature of the conversions between mechanical energy and internal
energy, which is then addressed here in details.
Section 2 provides the general theory.
Section 3 applies the result to elucidating the thermodynamics
of a Boussinesq fluid with a linear equation of state that has
been widely used recently in numerical study of turbulent mixing,
as well as in the context of horizontal convection. Section 4
discusses the case of the Boussinesq primitive equations 
currently used in numerical ocean models. Section 5 summarises
and discusses some implications of the results.

\section{Thermodynamically consistent and inconsistent
Boussinesq/anelastic models}
\label{main}

\subsection{Some specific examples of inconsistent Boussinesq models}

 To help set up the context and the motivation for the present work,
it is useful to provide some specific examples of energetically and
thermodynamically inconsistent Boussinesq approximations, which have
played and often continue to play a key role in the theoretical and
numerical study of many fluid flows of interest in traditional and
geophysical fluid dynamics.

\subsubsection{Boussinesq fluid with a linear equation of state}

 The first Boussinesq model of interest is one that is meant to approximate
fluids with an equation of state close to linear, retaining all physical
processes apart from sound waves from the largest scales down to the molecular
diffusive and dissipative scales. Its governing equations are:
\begin{equation}
    \frac{D{\bf v}}{Dt} + \frac{1}{\rho_0} \nabla \delta P = 
   - \frac{g_0(\rho-\rho_0)}{\rho_0} {\bf k} + \nu \nabla^2 {\bf v} ,
   \label{velocity_model1}
\end{equation}
\begin{equation}
        \nabla \cdot {\bf v} = 0 ,
    \label{continuity_model1}
\end{equation}
\begin{equation}
     \frac{DT}{Dt} = \kappa \nabla^2 T ,
     \label{temperature_model1}
\end{equation}
\begin{equation}
     \rho = \rho_0 \left [ 1- \alpha (T-T_0) \right ] ,
     \label{eos_model1}
\end{equation}
where ${\bf v}=(u,v,w)$ is the three-dimensional velocity field,
$\delta P=P-P_0$ is the pressure anomaly defined relative to the
reference Boussinesq pressure $P_0=-\rho_0 g_0 z$, 
$\rho$ is the density, $T$ is the temperature,
$\rho_0$ and $T_0$ are reference constant density and temperature,
${\bf k}$ is the unit normal vector pointing upwards in the direction
opposite to gravity, and $g_0$ is a
nominal value of the acceleration of gravity. Although such a model
is neither energetically nor thermodynamically consistent, as made
clear in this paper, it has
nevertheless been extensively used in recent theoretical 
discussion of the energetics of horizontal convection, e.g., 
\cite{Paparella2002}, \cite{Wang2005} and 
\cite{Winters2009}, as well as in discussing the energetics of
turbulent mixing in stratified fluids by \cite{Winters1995}.
Moreover, such a model also forms the basis for numerous direct 
numerical simulations of stratified turbulence, in the sense that
in such studies, both $\kappa$ and $\nu$ are usually interpreted as 
representing the molecular values of diffusivity and viscosity
respectively. From a thermodynamic viewpoint, the physical 
meaning of the temperature $T$ is ambiguous, since it is conserved
for adiabatic motions as if it were potential temperature, while
also being homogenised by molecular diffusion as if it were 
in-situ temperature. Regarding pressure, it can be written as the
sum of a purely hydrostatic component plus a perturbation $P'$ that
can be regarded as the Lagrange multiplier associated with 
the incompressibility condition 
$\nabla \cdot {\bf v}=0$, and hence differs from the thermodynamic
pressure.

\subsubsection{Coarse-grained Boussinesq model with parameterised turbulent
fluxes}

 A second important class of Boussinesq model is that associated with
the kind of hydrostatic primitive equations model that has formed the
basis for most numerical ocean models currently in used for climate 
change studies. An early formulation, following that introduced by 
\cite{Bryan1969}, is the following: 
\begin{equation}
    \frac{D{\bf u}}{Dt} + f {\bf k} \times {\bf u}
   + \frac{1}{\rho_0} \nabla_h \delta P = A_H \nabla_h^2 {\bf u} + 
   A_v \frac{\partial^2 {\bf u}}{\partial z^2} ,
   \label{velocity_ogcm}
\end{equation}
\begin{equation}
       \frac{1}{\rho_0} \frac{\partial \delta P}{\partial z} = 
       - g_0 \left ( \frac{\rho - \rho_0}{\rho_0} \right ),
      \label{hydrostatic_ogcm}
\end{equation}
\begin{equation}
      \nabla_h \cdot {\bf u} + \frac{\partial w}{\partial z} = 0 ,
     \label{continuity_ogcm}
\end{equation}
\begin{equation}
        \frac{D\theta}{Dt} = K_H \nabla_h^2 \theta + 
       K_v \frac{\partial^2 \theta}{\partial z^2}  ,
       \label{temperature_ogcm}
\end{equation}
\begin{equation}
        \frac{DS}{Dt} = K_H \nabla_h^2 S + 
        K_v \frac{\partial^2 S}{\partial z^2} ,
       \label{salinity_ogcm}
\end{equation}
\begin{equation}
      \rho = \rho(\theta,S,P) ,
      \label{eos_ogcm}
\end{equation}
where ${\bf u}=(u,v)$ is the horizontal velocity field, $w$ is the
vertical velocity, $\nabla_h$ is the horizontal nabla operator, 
$A_H$ and $A_V$ represents horizontal and vertical turbulent eddy
viscosities, while $K_H$ and $K_V$ represent horizontal and vertical
turbulent eddy diffusivities. Here, the full nonlinear equation of
state for seawater is used, and is usually formulated in terms of salinity $S$,
potential temperature $\theta$ and pressure $P$. Regarding the
latter, it is worth pointing out that there is currently much debate
about whether the Boussinesq pressure $P_0=-\rho_0 g_0 z$ or the
full hydrostatic pressure $P$ should be used in the equation of 
state, see \cite{Shchepetkin2011} for a discussion of this point.
Note that turbulent closure schemes in modern versions of primitive
equations models are in general significantly more sophisticated than
in the above model, see \cite{Griffies2004} for
a detailed discussion of current numerical ocean model formulations.
Until very recently, it was generally thought impossible for Boussinesq
primitive equations models such as the one above to admit a closed
energy budget when using a nonlinear equation of state; as a result,
numerical and empirical considerations must have been key in controlling 
the energetics of current numerical implementations of numerical ocean
models and hence their numerical stability.

\subsection{Dynamical/thermodynamic coupling in Boussinesq/anelastic models}

The first step toward constructing an energetically and thermodynamically
consistent Boussinesq/anelastic approximation for an arbitrary nonlinear
equation of state is to clarify the nature of the coupling between 
the dynamics and thermodynamics in a Boussinesq/anelastic fluid. In a real
fluid, such a coupling is achieved via the work of expansion/contraction
which occurs through the term $PD\upsilon/Dt$, where $P$ is the total 
pressure, and $\upsilon$ the specific volume. To fix ideas, we examine 
the issue in the context of the following set of equations, which is based
on that previously considered by \cite{Ingersoll2005} and \cite{Pauluis2008}:
\begin{equation}
    \frac{D{\bf v}}{Dt} + \nabla \cdot \left ( \frac{\delta P}{\rho_0} 
    \right ) = - \left ( \frac{\rho-\rho_0}{\rho_0} \right )g_0 \nabla Z
   + \frac{1}{\rho_0} \nabla \cdot {\bf S} ,
    \label{v_equation}
\end{equation}
\begin{equation}
         \nabla \cdot (\rho_0 {\bf v}) = 0 ,
     \label{continuity_equation}
\end{equation}
\begin{equation}
       \frac{D\eta}{Dt} = \dot{\eta} = \frac{\dot{q}}{T}
        = -\frac{1}{\rho_0} \nabla \cdot ( \rho_0 {\bf F}_{\eta})
       + \dot{\eta}_{irr} ,
       \label{eta_equation}
\end{equation}
\begin{equation}
       \frac{DS}{Dt} = \dot{S} = -\frac{1}{\rho_0} \nabla \cdot
       (\rho_0 {\bf F}_S) ,
       \label{s_equation}
\end{equation}
\begin{equation}
        \rho = \rho ( S,\eta,P_0) .
        \label{eos}
\end{equation}
The above system of equations is sufficiently general that it also
includes the usual Boussinesq approximation $\rho_0 = {\rm constant}$
as a particular case, as is easily verified. For that reason, 
Eqs. (\ref{v_equation}-\ref{eos}) will be hereafter referred to as the 
{\em Boussinesq-Anelastic} system, 
or {\em BA} system for short, where $\eta$ is the specific entropy,
$S$ is salinity, $g_0$ is a constant acceleration of gravity, $Z$ is
the geopotential height, so that $g_0 Z= \Phi$ is the geopotential, 
and ${\bf S}$ is the stress tensor. The terms $\dot{\eta}=\dot{q}/T$
and $\dot{S}$ are used as short-hand to denote the diabatic effects
affecting $\eta$ and $S$.
These are further specified in terms of an entropy flux ${\bf F}_{\eta}$
and irreversible entropy production term $\dot{\eta}_{irr}>0$ in Eq.
(\ref{eta_equation}), which is the expression of the second law of
thermodynamics, as well as in terms of a salt flux in eq. 
(\ref{s_equation}), so that salt is assumed to be a conservative 
quantity.
 It is important to remark at this stage that the definition of 
buoyancy $b=-g_0(\rho-\rho_0)/\rho_0$ is the one that is the most
commonly encountered in the literature. This differs, however, from
the form $b=g_0(\upsilon-\upsilon_0)/\upsilon_0$ assumed by 
\cite{Pauluis2008} and \cite{Young2010}. As shown in this paper, 
it turns out that the form used by \cite{Pauluis2008} and 
\cite{Young2010} is the one that should be used in practice, and
the one that is the most energetically and thermodynamically consistent.

\par

 Before proceeding, it seems important to point out that 
although \cite{Pauluis2008}'s derivation of the above BA system
initially assumes the reference pressure $P_0$ to satisfy
the classical hydrostatic balance 
$\rho_0^{-1} \partial P_0/\partial z=-g_0 \partial Z/\partial z$, 
this is no longer the case in the final energy conserving form of the
equations. Indeed, in absence of fluid motion, setting $P$ and
$\rho$ to zero in Eq. (\ref{v_equation}) shows that $P_0$ and
$\rho_0$ must actually be solution of:
\begin{equation}
     \frac{\partial}{\partial z} \left ( \frac{P_0}{\rho_0} 
     \right ) = - g_0
    \frac{\partial Z}{\partial z} ,
    \label{hydro_balance_error}
\end{equation}
which can be immediately integrated as: $P_0 = P_a -\rho_0 g Z$, 
where $P_a$ is the assumed constant and spatially uniform atmospheric
pressure at $Z=0$. This form is identical to the 
Boussinesq reference pressure, except for a non-constant $\rho_0$.
This in turn implies that:
\begin{equation}
     \rho_0 \frac{D}{Dt} \frac{P_0}{\rho_0} = 
    \frac{DP_0}{Dt} + \rho_0 P_0 \frac{D\upsilon_0}{Dt} = \nabla \cdot ( P_0
   {\bf v} ) = 
    -\rho_0 g_0 \frac{DZ}{Dt}  = \frac{DP_{0h}}{Dt} ,
   \label{pressure_error}
\end{equation}
where $P_{0h}$ is the reference pressure in hydrostatic equilibrium
with $\rho_0$, which shows that the relative error in $P_0$ as 
compared to $P_{0h}$, i.e., $\delta P_0/P_0 = O( \delta \rho_0/\rho_0)$,
and is therefore small by assumption.
Therefore, even though the anelastic approximation attempts to
relax the incompressibility constraint of the Boussinesq approximation,
\cite{Pauluis2008}'s energy conserving form does so at the price of
distorting the hydrostatic modes of motions, so that it would be of interest
to study the dynamical consequences of the BA system in more details,
for instance by following \cite{Davies2003}'s approach. The issue of
whether the distortion of the hydrostatic modes, 
which is not discussed in \cite{Pauluis2008}, could invalidate the
approach is beyond the scope of this paper, which is primarily concerned
with thermodynamic and energetic consistency issues.

\par

 In order to discuss the energetics of the BA system, our starting point
is the evolution equation for the kinetic energy, obtained in the usual
way by multiplying Eq. (\ref{v_equation}) by $\rho_0 {\bf v}$, which 
after some manipulation can be written as follows:
\begin{equation}
 \rho_0 \frac{D}{Dt} \frac{{\bf v}^2}{2} + \nabla \cdot (
  \delta P {\bf v} -\rho_0 {\bf F}_{ke})
   = \rho_0 b \frac{DZ}{Dt} - \rho_0 \varepsilon_K ,
  \label{ke_equation}
\end{equation}
where the work against the stress tensor has been written as
${\bf v} \cdot \nabla {\bf S} = \nabla \cdot ( {\bf v}\cdot
{\bf S} ) - \rho_0 \epsilon_K = \nabla \cdot (\rho_0 {\bf F}_{ke}) -
\rho_0 \varepsilon_K$ as the difference between the divergence of a flux
term minus the viscous dissipation term. For instance, in the special
case where $\rho_0^{-1} \nabla \cdot {\bf S} = \nu \nabla^2 {\bf v}$, then
${\bf F}_{ke} = \nu \nabla {\bf v}^2/2$ and $\varepsilon_K
= \nu( \| \nabla u \|^2 + \| \nabla v \|^2 + \| \nabla w \|^2)$.
Over the past decade, a 
number of studies by
\cite{Ingersoll2005}, \cite{Vallis2006}, \cite{Pauluis2008}, 
\cite{Young2010}, \cite{Nycander2010}, and \cite{McIntyre2010}
all followed a similar approach in seeking to establish the energetic
consistency of the above BA system, with the studies differing only
in minor details and in the particular set of equations considered.
Specifically, all these studies approached the problem by introducing
the following function:
\begin{equation}
      h^{\ddag}(S,\eta,Z) = h_0^{\ddag}(S,\eta) - \int_{Z_0}^{Z}
      b(S,\eta,Z'){\rm d}Z'
      \label{hdag_definition}
\end{equation}
obtained by vertically integrating the buoyancy $b$ regarded 
as a function of $Z$ and of the adiabatically conserved variables,
so that:
\begin{equation}
    \frac{Dh^{\ddag}}{Dt} = - b \frac{DZ}{Dt} 
    + C_S \dot{S} + C_{\eta} \dot{\eta} ,
    \label{h_equation}
\end{equation}
where
\begin{equation}
        C_S = \frac{\partial h_0^{\ddag}}{\partial S}
   - \int_{Z_0}^Z \frac{\partial h^{\ddag}}{\partial S}dZ' , \qquad
        C_{\eta} = \frac{\partial h_0^{\ddag}}{\partial \eta}
   - \int_{Z_0}^{Z} \frac{\partial h^{\ddag}}{\partial \eta} {\rm d}Z' .
\end{equation}
Then, summing Eqs. (\ref{ke_equation}) and (\ref{h_equation}) yields 
\begin{equation}
    \rho_0 \frac{D}{Dt} \left ( \frac{{\bf v}^2}{2} + h^{\ddag} \right )
  + \nabla \cdot \left ( \delta P {\bf v} - \rho_0 {\bf F}_{ke} \right )
   = \underbrace{\rho_0 \left ( C_S \dot{S} + C_{\eta} \dot{\eta} \right ) -
   \rho_0 \varepsilon_K}_{{\cal R}} .
  \label{energy_conservation_previous}
\end{equation}
Because the right-hand side ${\cal R}$ of 
Eq. (\ref{energy_conservation_previous}) obviously vanishes for adiabatic
and inviscid motions, it may be concluded that the quantity 
$\rho_0 ({\bf v}^2/2+h^{\ddag})$ should be regarded as the relevant energy
quantity being conserved in absence of diabatic and viscous effects.
Note that the energy quantity thus constructed is non-unique, since it
involves the arbitrary constant of integration $h_0^{\ddag}(S,\eta)$ as
discussed by \cite{Pauluis2008}. In the context of the Boussinesq 
approximation, both \cite{Young2010} and \cite{Nycander2010} 
chose $Z_0=h_0^{\ddag}=0$ and referred to $h^{\ddag}$ as the 
``dynamic enthalpy'' and ``effective potential energy'' respectively.
  
\par

 The above proof is incomplete, however, because establishing the 
energy consistency of the BA system actually requires: 
(1) Identifying the quantity playing the role of the total energy;
 (2) Demonstrating that the total energy thus identified is a conservative quantity
for both diabatic and adiabatic motions.
By contrast, all what the above arguments manage to establish is 
that the ``adiabatic'' energy ${\bf v}^2+h^{\ddag}$ is a conservative
quantity for adiabatic motions only. When diabatic effects are retained,
this adiabatic quantity is obviously created or destroyed owing to the
r.h.s. of Eq. (\ref{energy_conservation_previous}) becoming non-zero,
and hence no longer conservative. In that case, the principle of energy
conservation calls 
for the existence of an additional energy quantity
$h^{\ast}$ obeying an equation of the type:
\begin{equation}
    \rho_0 \frac{Dh^{\ast}}{Dt} + \nabla \cdot (\rho_0 {\bf F}_{h^{\ast}}) = 
  -{\cal R},
  \label{hstar_equation}
\end{equation}
involving some energy flux $\rho_0 {\bf F}_h^{\ast}$ to be determined, as 
upon summation with Eq. (\ref{energy_conservation_previous}), 
the following conservation equation is obtained
\begin{equation}
   \rho_0 \frac{D}{Dt} \left ( \frac{{\bf v}^2}{2} + h^{\ddag} + 
    h^{\ast} \right ) + \nabla \cdot ( \delta P {\bf v} - \rho_0 {\bf F}_{ke}
    + \rho_0 {\bf F}_h^{\ast} ) = 0 ,
    \label{total_energy_conservation}
\end{equation}
which now states that the energy quantity 
${\bf v}^2 + h^{\ddag} + h^{\ast}$ is conservative for both diabatic and
adiabatic motions, and hence the natural candidate to play the role of
the total energy in the BA system.

\par

  Physically, the above approach casts the discussion of energetics in
terms of the interactions between the adiabatic energy 
${\bf v}^2/2+h^{\ddag}$ and the diabatic energy $h^{\ast}$. While 
legitimate and sometimes useful (as attested by the papers by 
\cite{Nycander2010} and \cite{McIntyre2010} for instance), this
approach cannot be regarded as entirely satisfactory, since: 
1) the adiabatic and diabatic energies are fundamentally ill defined
quantities, given that equally valid alternatives can be obtained by adding
and subtracting any arbitrary function of $S$ and $\eta$ from the former
and to the later respectively; 
2) the approach does not appear to be naturally linked to the 
classical description in terms of interactions
between kinetic energy, gravitational potential energy and internal
energy. For instance, can we regard the sum $h^{\ddag}+h^{\ast}$ 
as equal to the sum of gravitational potential energy and internal 
energy, as for actual fully compressible fluids? How do energy 
conversions in the BA system compare with that of a fully compressible
fluid?

\par

 To address the above questions, a fundamentally different approach is
clearly needed. If we are to understand the coupling between mechanical
energy and internal energy in the BA system, it seems necessary to first 
identify what is the relevant expression for the mechanical energy, and
then look for its evolution equation in order to determine how it couples
with the internal energy. The most natural definition for the later 
appears to be:
$\rho_0 E_m = \rho_0 [{\bf v}^2/2 - b (Z-Z_0)] =
\rho_0 {\bf v}^2/2 + (\rho-\rho_0)g_0(Z-Z_0)$ 
(for some reference geopotential height $Z_0$), for which an evolution
equation can be obtained from the kinetic energy
equation (\ref{ke_equation}) as follows:
\begin{equation}
   \rho_0 \frac{D}{Dt} \left [ \frac{{\bf v}^2}{2} -
   b(Z-Z_0) \right ] + 
  \nabla \cdot (\delta P {\bf v} - \rho_0 {\bf F}_{ke}) 
   = -\rho_0 (Z-Z_0) \frac{Db}{Dt} 
   - \rho_0 \varepsilon_K,
  \label{anelastic_em1}
\end{equation}
or equivalently, in conservative form: 
\begin{equation}
   \frac{\partial (\rho_0 E_m)}{\partial t} + \nabla \cdot 
   \left [ \rho_0 \left ( E_m + \frac{\delta P}{\rho_0} \right )
  {\bf v} - \rho_0 {\bf F}_{ke} \right ] = 
    -\rho_0 (Z-Z_0) \frac{Db}{Dt} -\rho_0 \varepsilon_K .
  \label{anelastic_em2}
\end{equation}
For comparison, the corresponding evolution equation for the
mechanical energy in a fully compressible fluid takes the form:
\begin{equation}
   \rho \frac{D}{Dt} \left [ \frac{{\bf v}^2}{2} + g_0 (Z-Z_0) 
  \right ] + \nabla \cdot \left ( P {\bf v} - \rho {\bf F}_{ke} \right )
  = \rho P \frac{D\upsilon}{Dt} - \rho \varepsilon_K .
  \label{compressible_em}
\end{equation}
In a compressible fluid, mechanical energy and internal energy
are coupled in mainly two ways: (i) via a reversible conversion
between internal energy and kinetic energy accomplished by the work of
expansion/contraction $P D\upsilon/Dt$; (ii) the irreversible dissipation of kinetic
energy into heat by viscous processes $\varepsilon_K$, as seen in the
r.h.s. of Eq. (\ref{compressible_em}). Since viscous dissipation is obviously
present in both Eqs. (\ref{anelastic_em1}) and 
(\ref{compressible_em}), the only question that needs addressing is whether
the following equivalence can be established?
\begin{equation}
         - \rho_0 (Z-Z_0) \frac{Db}{Dt} \qquad \leftrightarrow \qquad
         \rho P \frac{D\upsilon}{Dt} \qquad ?
          \label{correspondance}
\end{equation}
To see that this is indeed the case, we expand the pressure and
specific volume as follows:
\begin{equation}
      P = P_a - \rho_0 g_0 Z + \cdots , \qquad
     \upsilon = \frac{1}{\rho_0} - \frac{\rho-\rho_0}{\rho_0^2} + \cdots
     \label{boussinesq_expansion}
\end{equation}
which uses the underlying assumption in the Boussinesq and 
anelastic approximations that $(\rho-\rho_0)/\rho_0 \ll 1$.
This implies in turn that at leading order, the work of 
expansion/contraction can be approximated by:
$$
      \rho P \frac{D\upsilon}{Dt} \approx
       \rho_0 (P_a - \rho_0 g_0 Z) 
     \frac{D}{Dt} \left (\frac{1}{\rho_0} -
    \frac{\rho-\rho_0}{\rho_0^2} \right )
$$
\begin{equation}
    = - \rho_0 (Z-Z_0) \frac{Db}{Dt} + \rho_0 (P_a-\rho_0 g_0 Z)
     \frac{D\upsilon_0}{Dt} .
   \label{pv_result}
\end{equation}
Eq. (\ref{pv_result}) shows that the equivalence Eq. (\ref{correspondance})
is exact in the Boussinesq case $\rho_0 = {\rm constant}$, but only approximate
in the anelastic case because of the term proportional to $D\upsilon_0/Dt$.
This suggests, therefore, that the term $\delta W_{ba} = (Z-Z_0){\rm d}b$ is
the natural counterpart of the compressible work $\delta W = - P{\rm d}\upsilon$
for a compressible fluid. The possibility to identify $\delta W_{ba}$ with $\delta W$
is a key result of this paper, for it points to a natural way
to construct the whole range of known thermodynamic potentials for a BA 
fluid, as shown in the next subsection \ref{thermo_ba}. 
To be fair, the identification of $\delta W_{ba}$
with $\delta W$ is not entirely new, since it was pointed out in earlier
studies such as \cite{Winters1995},
\cite{Wang2005} and \cite{Nycander2007} for the kind of Boussinesq 
fluid with a linear equation of state discussed in Section \ref{linear_eos}.
As far as we are aware, however, its 
consequences for constructing thermodynamically consistent Boussinesq
and anelastic approximations do not appear to have been realized until now.
Physically, Eq. (\ref{correspondance}) can be interpreted as expressing the
fact that in a BA fluid, the compressible work manifests itself
through the apparent changes in gravitational potential energy due to the
apparent changes in the mass element ${\rm d} m = \rho{\rm d}V$, which
is possible since a BA fluid conserves the reference
mass element ${\rm d}m_0 = \rho_0 {\rm d}V$ rather than ${\rm d}m$.

\subsection{Thermodynamics of Boussinesq and Anelastic binary fluids}
\label{thermo_ba}

  Having established in the previous section that the energetics of
both the Boussinesq and anelastic equations possess a term that can be
identified as a conversion between internal energy and mechanical 
energy, it then becomes relatively straightforward to show that this
can be used to construct energetically and thermodynamically consistent
BA system of equations. To that end, let us first recall that
a basic tenet of classical equilibrium thermodynamics is that the 
specific internal energy $e=e(\eta,S,\upsilon)$ of any fluid in local
thermodynamic equilibrium (LTE) can be regarded as 
a function of state whose value is independent of the thermodynamic
path followed, where $\eta$ is the specific entropy, $S$ is salinity
(the argument would also work if $S$ was the total mixing ratio, as
for the case of moist air considered by \cite{Pauluis2008}), 
and $\upsilon$ is the specific volume. As is well known, this implies
that the reversible work
transfer $\delta W=-P{\rm d}\upsilon$ and generalised heat transfer
$\delta Q = T {\rm d}\eta + \mu {\rm d}S$ entering the total differential
of $e$, viz.,
\begin{equation}
     {\rm d} e = \delta Q + \delta W = T {\rm d}\eta + \mu {\rm d}S
  - P {\rm d}\upsilon ,
   \label{fundamental_relation}
\end{equation}
cannot be specified independently of each other. In classical 
equilibrium thermodynamics, this interdependence is imposed by
the so-called Maxwell relationships, which simply express the
result that the cross-derivatives with respect to two 
different variables must be equal for twice continously differentiable
functions. In the present case, the assumption that $e$ is a function
of the thermodynamic state only implies that:
\begin{equation}
       T = \frac{\partial e}{\partial \eta}, \qquad
       \mu = \frac{\partial e}{\partial S}, \qquad
       P = -\frac{\partial e}{\partial \upsilon},
\end{equation}
which in turn implies the following three Maxwell relationships:
\begin{equation}
     \frac{\partial T}{\partial \upsilon} =
    \frac{\partial^2 e}{\partial \eta \partial \upsilon} =
    -\frac{\partial P}{\partial \eta},
  \qquad \frac{\partial T}{\partial S} =
    \frac{\partial^2 e}{\partial \eta \partial S} = 
   \frac{\partial \mu}{\partial \eta},
   \qquad
    \frac{\partial \mu}{\partial \upsilon} = \frac{\partial^2 e}
   {\partial S \partial \upsilon} =  -\frac{\partial P}{\partial S} .
\end{equation}
The above remarks suggest that the simplest way to ensure that the
BA system is energetically and thermodynamically consistent is to
ensure that the approximation to the generalised heat transfer, 
denoted here by $\delta Q_{ba} = T_{ba} {\rm d}\eta_{ba} + 
\mu_{ba} {\rm d} S_{ba}$, is similarly linked via 
relevant Maxwell relationships
to the approximation to the work transfer identified previously, viz.,
\begin{equation}
    \delta W_{ba} = (Z-Z_0){\rm d}b,
\end{equation}
where the subscript {\em ba} was used to denote the approximation
to the generalised `heat variables and functions' 
$T$, $\eta$, $\mu$ and $S$. If so, this would
in turn allows one to regard the following expression:
\begin{equation}
     {\rm d}e_{ba} = T_{ba} {\rm d}\eta_{ba} + \mu_{ba} {\rm d}S_{ba}
   + (Z-Z_0){\rm d}b,
   \label{eba}
\end{equation}
as the natural counterpart of the fundamental relation of thermodynamics
Eq. (\ref{fundamental_relation}),
and hence as the total differential of the relevant approximation to the
internal energy of the `BA fluid', for which the natural variables are 
entropy $\eta$, salinity $S$, and buoyancy $b$. In practice, however,
it is often more convenient to work with pressure $P$ as a dependent
variable rather than specific volume, as well as with temperature $T$ rather
than entropy $\eta$, which motivates the introduction of additional
thermodynamic potentials constructed from Eq. (\ref{fundamental_relation})
by means of the Legendre transform, e.g., \cite{Alberty2001}.
The most common 
thermodynamic potentials that are also the most relevant for the
present work are the specific enthalpy $h=e+p\upsilon$ and the
Gibbs free energy $g = e+p\upsilon-T\eta$, whose natural dependent
variables are $(\eta,S,P)$ and $(T,S,P)$ respectively. From Eq. 
(\ref{eba}), it is easy to convince oneself that the corresponding
approximations to $h$ and $g$ are given by:
$h_{ba} = e_{ba} - b (Z-Z_0)$ and $g_{ba} = e_{ba} - b (Z-Z_0) 
  - T_{ba} \eta_{ba}$, with the following total differentials:
\begin{equation}
    {\rm d}h_{ba} = d \left [ e_{ba} - b (Z-Z_0) \right ] = T_{ba}
  {\rm d}\eta_{ba} + \mu_{ba} {\rm d}S_{ba} - b {\rm d}Z
  \label{ba_enthalpy}
\end{equation}
\begin{equation}
    {\rm d}g_{ba} = d \left [ e_{ba} - b(Z-Z_0) - T_{ba} \eta_{ba} \right ]
   = - \eta_{ba} {\rm d}T_{ba} + \mu_{ba} {\rm d}S_{ba} - b {\rm d}Z .
  \label{ba_gibbs}
\end{equation}
Until now, the above considerations have remained rather formal, and
while they indicate that it is in principle possible to construct the
full range of known thermodynamic potentials for a BA fluid, they 
have not addressed the issue of how such potentials might be constructed
in practice. To simplify notations, we will usually refrain from using
the suffix $ba$ for the dependent variables in the following. We should
keep in mind, however, that all thermodynamic variables used in the
context of the Boussinesq/anelastic approximations are fundamentally
different, even if possibly only very slightly in numerical values, from
their actual counterparts.

\par

 In the previous approaches by \cite{Ingersoll2005}, \cite{Pauluis2008},
\cite{Young2010} and \cite{Nycander2010}, authors have generally assumed
the buoyancy $b=b(\eta,S,Z)$ to be given as a function of the dependent
variables considered, typically entropy, salinity and geopotential height $Z$.
From Eq. (\ref{ba_enthalpy}), this can be integrated with respect to $Z$,
viz.,
\begin{equation}
        h_{ba}(\eta,S,Z) = h_{ba}(\eta,S,Z_0) - \int_{Z_0}^Z 
        b(\eta,S,Z')\,{\rm d}Z',
\end{equation}
for some reference geopotential height $Z_0$, traditionally taken at
the ocean surface in the oceanic case,
but this only provides an expression for the specific enthalpy up to
an indeterminate function of entropy and salinity, which cannot be
specified without additional thermodynamic information about the fluid.

\par
  
 The above problem arises because the knowledge of density alone is
insufficient to specify all possible thermodynamic quantities; the
heat capacity, for instance, cannot be inferred from density. On the
other hand, many thermodynamic potentials have the property
of encapsulating all known thermodynamic
knowledge about a given fluid, as discussed by \cite{Callen1985}.
In the following, we show how to obtain the relevant approximations
to the Boussinesq/anelastic system directly from two such
thermodynamic potentials, the Gibbs function and the enthalpy.

\subsubsection{Deriving BA thermodynamic potentials from exact
thermodynamic potentials}

 Given the similarity of the approximate and exact differentials
for the internal energy, enthalpy, and Gibbs free energy, it is
natural to ask whether it is possible to compute the approximate
thermodynamic potentials from their exact counterparts, rather 
than by integrating the total differential of the approximate
thermodynamic potentials.
That this is indeed the case is shown first for the Gibbs free
energy, which is the thermodynamic potential currently
chosen to express all
the thermodynamic properties for seawater in the latest standard
recently adopted by UNESCO, see \cite{teos10}, as advocated 
earlier by \cite{Feistel2003} for instance. To that end, let us
introduce the following function:
\begin{equation}
      g_{ba} = g_{ba}(T,S,Z) = g(T,S,P_{0h}(Z)) + g_0(Z-Z_0)
     = \tilde{g} + g_0(Z-Z_0),
      \label{gba_trial}
\end{equation}
and verify that its total differential agrees with the above
differential relationship for the approximate Gibbs function,
where $g=g(T,S,P)$ is the exact Gibbs free energy expressed
in terms of its natural variables $T$, $S$ and $P$. Note that
the approximate pressure used in Eq. (\ref{gba_trial}) is the
pressure $P_{0h}$ in hydrostatic balance with $\rho_0$, not
the anelastic reference pressure $P_0=P_a - \rho_0 g_0 Z$.
From the result that ${\rm d}g  = - \eta {\rm d}T + \mu 
{\rm d}S + \upsilon {\rm d}P$, it follows that the total   
differential of $g_{ba}$ is given by:
\begin{equation}
     {\rm d}g_{ba} = -\tilde{\eta} {\rm d}T + \tilde{\mu} {\rm d}S +
     g_0 (1-\rho_0\tilde{\upsilon}) {\rm d}Z,  
     \label{gba_differential}
\end{equation}
where we used ${\rm d}P_{0h} = -\rho_0 g_0 {\rm d}Z$, and
where the tilde quantities are estimated for the reference pressure
$P_{0h}(Z)$ instead of the actual pressure, so that
$\tilde{\eta} = \eta(T,S,P_{0h}(Z))$, $\tilde{\mu}=\mu(T,S,P_{0h}(Z))$
and $\tilde{\upsilon} = \upsilon(T,S,P_{0h}(Z))$. 
Eq. (\ref{gba_differential}) is obviously consistent with 
Eq. (\ref{ba_gibbs}) provided that $\eta_{ba}$, $\mu_{ba}$ and $b$
be defined by:
\begin{equation}
     \eta_{ba} = \tilde{\eta}, \qquad \mu_{ba} = \tilde{\mu}, \qquad
     b = - g_0 (1-\rho_0 \tilde{\upsilon}) = - g_0 
     \left ( \frac{\tilde{\rho}-\rho_0}{\tilde{\rho}} \right ) .
    \label{main_results}
\end{equation}
The key result here is that while $\eta_{ba}$ and $\mu_{ba}$ are 
found to be identical to their natural
tilded expressions, the density $\rho$
initially used to defined the buoyancy $b=-g_0(\rho-\rho_0)/\rho_0$
is on the other hand found to differ from its natural tilded
expression $\tilde{\rho}$, as Eq. (\ref{main_results}) shows that:
\begin{equation}
       b = - g_0 \left ( \frac{\tilde{\rho}-\rho_0}{\tilde{\rho}} 
      \right ) = - g_0 \left ( \frac{\rho-\rho_0}{\rho_0} \right ),
\end{equation}
which implies that:
\begin{equation}
   \rho = \rho_0\left ( 1 + \frac{\tilde{\rho}-\rho_0}{\tilde{\rho}} 
  \right ) = \rho_0 \left ( 2 - \rho_0 \tilde{\upsilon} \right ) .
  \label{true_buoyancy}
\end{equation}
Interestingly, the definition of the buoyancy in Eq. (\ref{main_results})
is identical to the one recently proposed by \cite{Pauluis2008} and
\cite{Young2010}, but differs from the definition usually adopted in most
numerical ocean general circulation models as far as we know.

\par

 As seen above, the knowledge of $g_{ba}$ is sufficient by itself to
determine all possible thermodynamic potentials. Thus, the expressions
for the specific internal energy and enthalpy are given by:
\begin{equation}
    e_{ba}(S,T,Z) = g_{ba} + b(Z-Z_0) + T_{ba} \eta_{ba} = 
         \tilde{g} + (g_0+b)(Z-Z_0) + T \tilde{\eta} , 
\end{equation}
\begin{equation}
    h_{ba}(S,T,Z) = e_{ba} - b(Z-Z_0) = \tilde{g} + g_0(Z-Z_0)
      + T \tilde{\eta} ,
\end{equation}
which provide expressions for the specific internal energy and
enthalpy in terms of the Gibbs function natural variables $T$, $S$, 
and $Z$, rather than in terms of $e$ and $h$'s natural variables.
It is easily verified that the total differential for $e_{ba}$ and
$h_{ba}$ are given by:
\begin{equation}
    {\rm d}e_{ba} = T {\rm d} \tilde{\eta} + \tilde{\mu} {\rm d}S 
+ (Z-Z_0) {\rm d}b,
  \label{de_ba}
\end{equation}
\begin{equation}
    {\rm d}h_{ba} = T {\rm d}\tilde{\eta}
+ \tilde{\mu}{\rm d}S - b{\rm d}Z  ,
  \label{dh_ba}
\end{equation}
which can be checked to be in agreement with Eqs. (\ref{eba}) and
(\ref{ba_enthalpy}), thus completing the proof.

\par

  While the use of the Gibbs function is the most natural approach when the
thermodynamic properties are tabulated as functions of $T$, $S$, and
$P$, as is currently the case for seawater, e.g., \cite{teos10},
it is sometimes more convenient, depending on the particular 
situation considered, to work with different dependent variables and
hence with a different ``master'' thermodynamic potential. For a general
discussion of the different ways to compile thermodynamic data, the
interested reader is referred to \cite{Callen1985}. The specific enthalpy
$h=h(\eta,S,P)$, owing to the particularly important role it plays 
for describing and understanding the energetics of stratified fluids, warrants
a separate discussion and is therefore discussed next. To that end,
let us introduce the function $h_{ba}$:
\begin{equation}
  h_{ba} = h(\eta,S,P_{0h}(Z)) + g_0 (Z-Z_0) = \tilde{h} + g_0(Z-Z_0), 
\end{equation}
and verify that it is the relevant Boussinesq/anelastic approximation
to the specific enthalpy. This is done by taking the total differential of $h_{ba}$, viz.,
\begin{equation}
  {\rm d}h_{ba} = \tilde{T} {\rm d}\eta + \tilde{\mu} {\rm d}S
    + g_0(1 - \rho_0 \tilde{\upsilon}) {\rm d}Z
   = \tilde{T} {\rm d}\eta + \tilde{\mu} {\rm d}S - b {\rm d}Z ,
\end{equation}
and checking that it agrees with Eq. (\ref{ba_enthalpy}). QED.
The corresponding expression for the internal energy
$e_{ba} = h_{ba} + b (Z-Z_0)$ can be written as:
\begin{equation}
    e_{ba} = \tilde{h} + (g_0+b)(Z-Z_0) .
\end{equation}
Its total differential is:
\begin{equation}
   {\rm d}e_{ba} = \tilde{T} {\rm d}\eta + \tilde{\mu} {\rm d}S
   -\rho_0 g_0 \tilde{\upsilon} {\rm d}Z + (b+g_0) {\rm d}Z 
     + (Z-Z_0) {\rm d}b = \tilde{T}{\rm d}\eta + \tilde{\mu}
  {\rm d}S + (Z-Z_0) {\rm d}b ,
\end{equation}
which is again consistent with Eq. (\ref{eba}).

\par

 As a final remark, let us mention that while the above derivations 
 demonstrate that the BA thermodynamic 
potentials can be constructed from the knowledge of the exact 
Gibbs function or enthalpy, integration of the Maxwell relationships
will in general be needed to construct the BA thermodynamic potentials
when an idealised equation of state is assumed. Sections \ref{linear_eos}
and \ref{nonlinear_eos} illustrate in details the two different possible cases.

\subsection{Consequence for energy conservation and
``heat''-related quantities}

  The above results make it possible to write down the evolution equations
for the specific internal energy and enthalpy by combining Eqs. 
(\ref{de_ba}) and (\ref{dh_ba}) with the evolution equations 
Eqs. (\ref{eta_equation}) and (\ref{s_equation}) for $\eta$ and $S$
respectively, leading to:
\begin{equation}
    \rho_0 \frac{De_{ba}}{Dt} = \rho_0 [ \dot{q} + \tilde{\mu} \dot{S} ]
     + \rho_0 (Z-Z_0) \frac{Db}{Dt} ,
    \label{eba_evolution}
\end{equation}
\begin{equation}
     \rho_0 \frac{Dh_{ba}}{Dt} = \rho_0 \left [ \dot{q} + \tilde{\mu}
    \dot{S} \right ] - \rho_0 b \frac{DZ}{Dt} .
    \label{hba_evolution}
\end{equation}
Interestingly, taking the difference between these two equations
yields the evolution equation for the gravitational potential energy
$b(Z-Z_0)$:
\begin{equation}
          \rho_0 \frac{D}{Dt} \left [ b(Z-Z_0) \right ] = 
          \rho_0 \frac{D(h_{ba}-e_{ba})}{Dt} ,
\end{equation}
which shows that for a BA fluid, the gravitational potential energy
can be regarded as a thermodynamic property of the fluid, since
it is entirely determined from the knowledge of $S$, $T$, and $Z$.
There are therefore two main ways to describe the coupling between
dynamics and thermodynamics in a BA fluid, the first one focusing 
on the coupling between internal energy and the total mechanical
energy (the sum of kinetic energy and gravitational potential energy), 
and the other between the kinetic energy and the total 
potential energy (the sum of internal energy and gravitational 
potential energy, which is equal to the enthalpy here).
By combining either one of Eqs. (\ref{eba_evolution}) or 
(\ref{hba_evolution}) with the mechanical energy equation
Eq. (\ref{anelastic_em1}), the two following equivalent evolution
equations for the total energy $E_t = 
{\bf v}^2/2 + e_{ba} - b (Z-Z_0) = {\bf v}^2 + h_{ba}$
are obtained:
\begin{equation}
    \rho_0 \frac{D}{Dt} \left [ \frac{{\bf v}^2}{2} - b (Z-Z_0) 
     + e_{ba} \right ] + \nabla \cdot [ \delta P {\bf v} - 
    \rho_0 {\bf F}_{ke} ] = \rho_0 \left [ \dot{q} + \tilde{\mu} \dot{S} - 
    \varepsilon_K \right ] ,
   \label{total_energy}
\end{equation}
\begin{equation}
    \rho_0 \frac{D}{Dt} \left [ \frac{{\bf v}^2}{2} + h_{ba} \right ]
   + \nabla \cdot \left [ \delta P {\bf v} - 
    \rho_0 {\bf F}_{ke} \right ] = \rho_0 
    \left [ \dot{q} + \tilde{\mu} \dot{S} - \varepsilon_K \right ] .
   \label{total_energy_2}
\end{equation}
Now, in order for the BA system to be energetically consistent, it 
remains to verify that Eqs. (\ref{total_energy}) and (\ref{total_energy_2})
are consistent with the principle of energy conservation. This is easily
shown to be the case only if the right-hand side of Eq. (\ref{total_energy})
is the divergence of some energy flux ${\bf F}_q$, i.e.,
\begin{equation}
      \rho_0 (\dot{q} + \mu \dot{S} - \varepsilon_K) = -
     \nabla \cdot ( \rho_0 {\bf F}_q ) .
     \label{energy_flux}
\end{equation}
As we show now, this imposes a constraint on the form of the
entropy flux ${\bf F}_{\eta}$ and irreversible entropy production
$\dot{\eta}_{irr}$ in Eq. (\ref{eta_equation}).
Assuming salt to be conservative quantity, and hence such that
$\rho_0 \dot{S} = - \nabla \cdot (\rho_0 {\bf F}_S)$ for some salt
flux ${\bf F}_S$, implies for the evolution equation of specific entropy:
$$
    \rho_0 \frac{D\eta_{ba}}{Dt} = \frac{\rho_0 \dot{q}}{T} 
   = \frac{\rho_0 \varepsilon_K + \mu \nabla \cdot (\rho_0 {\bf F}_S)
    - \nabla \cdot (\rho_0 {\bf F}_q)}{T}
$$
\begin{equation}
     = \frac{\rho_0 \varepsilon_K}{T} 
    - \nabla \cdot \left \{ \frac{\rho_0 ( {\bf F}_q - \tilde{\mu}
   {\bf F}_S)}{T} \right \} + \rho_0 \left [ 
  {\bf F}_q \cdot \nabla \left ( \frac{1}{T} \right ) - 
  {\bf F}_S \cdot \nabla \left ( \frac{\tilde{\mu}}{T} \right ) 
  \right ] .
\end{equation}
The latter expression establishes that the entropy flux ${\bf F}_{\eta}$
and irreversible entropy production $\dot{\eta}_{irr}$ in 
Eq. (\ref{eta_equation}) must be related to the salt flux ${\bf F}_S$ and
enthalpy/internal energy flux ${\bf F}_q$ by:
\begin{equation}
         {\bf F}_{\eta} = \frac{{\bf F}_q - \tilde{\mu}{\bf F}_S}{T},
\end{equation}
\begin{equation}
         \dot{\eta}_{irr} = \frac{\varepsilon_K}{T} 
         + {\bf F}_q \cdot \nabla \left ( \frac{1}{T} \right ) 
         - {\bf F}_S \cdot \nabla \left ( \frac{\tilde{\mu}}{T} \right )  .
\end{equation}
In non-equilibrium thermodynamics, this is usually the point
at which the second law of thermodynamics is 
then invoked to further constrain ${\bf F}_q$ and ${\bf F}_S$, 
in order to guarantee that $\dot{\eta}_{irr} \ge 0$, e.g., see 
\cite{degroot1962}. Going back to the evolution equation for internal
energy and enthalpy, note that Eq. (\ref{energy_flux}) implies:
\begin{equation}
    \rho_0 \frac{De_{ba}}{Dt} = - \nabla \cdot (\rho_0 {\bf F}_q) 
    + \rho_0 \varepsilon_K + \rho_0 (Z-Z_0) \frac{Db}{Dt} ,
\end{equation}
\begin{equation}
    \rho_0 \frac{Dh_{ba}}{Dt} = -\nabla \cdot ( \rho_0 {\bf F}_q)
  + \rho_0 \varepsilon_K - \rho_0 b \frac{DZ}{Dt} ,
\end{equation}
so that ${\bf F}_q$ appears as the diffusive flux of internal energy
or enthalpy. Both equations can be regarded as equivalent forms of
the first law of thermodynamics.

\subsection{Alternative forms of the first law of thermodynamics}

 We now investigate the consequences of the above results for the
determination of various temperature variables. Let us first note that
from the differential of the Gibbs function:
$$
    {\rm d}g_{ba} = - \tilde{\eta} {\rm d}T + \tilde{\mu}{\rm d}S -b {\rm d}Z ,
$$
the Maxwell relationships provide the following partial derivatives for
the entropy:
$$
    \frac{\partial \tilde{\eta}}{\partial S} = 
- \frac{\partial \tilde{\mu}}{\partial T}, \qquad
    \frac{\partial \tilde{\eta}}{\partial Z} = \frac{\partial b}{\partial T} = \rho_0 g_0 
    \frac{\partial \tilde{\upsilon}}{\partial T} = g_0 \tilde{\alpha},
$$
by defining $\tilde{\alpha} = \rho_0 \partial \tilde{\upsilon}/\partial T$ as 
the relevant definition of the thermal expansion coefficient.
Using the well known result that $\partial \eta/\partial T = c_p {\rm d}T/T$,
this implies that we can write:
\begin{equation}
         {\rm d}\eta = \frac{\tilde{c}_p}{T} {\rm d}T - \frac{\partial \tilde{\mu}}{\partial T}
         {\rm d}S + g_0 \tilde{\alpha} {\rm d}Z
\end{equation}
As a result, it follows that the temperature equation can be written as:
\begin{equation}
    \frac{DT}{Dt} = \frac{T}{\tilde{c}_p} \frac{D\eta}{Dt} 
  + \frac{T}{\tilde{c}_p} \frac{\partial \tilde{\mu}}{\partial T} \frac{DS}{Dt}
     - \frac{\tilde{\alpha} g_0 T}{\tilde{c}_p} \frac{DZ}{Dt} .
\end{equation}
In absence of salinity, this equation takes the simpler form:
\begin{equation}
      \frac{DT}{Dt} = 
      \frac{\nabla \cdot (\kappa \tilde{c}_p \nabla T)}{\tilde{c}_p}
   + \frac{\varepsilon_K}{\tilde{c}_p} - \frac{\alpha g_0 T}{\tilde{c}_p}
   \frac{DZ}{Dt} ,
   \label{consistent_temperature}
\end{equation}
by assuming the diffusive heating to be given by the classical Fourier
law $\nabla \cdot ( \rho_0 {\bf F}_q ) = -\nabla \cdot ( \kappa \rho_0 
\tilde{c}_p \nabla T )$.
This shows that the evolution equation for in-situ temperature in
general possesses: a) a term related to molecular diffusion, that
is general not-conservative, i.e., it cannot be expressed as the
divergence of a flux because $c_p$ is not constant; b) it usually
incorporate the Joule heating due to viscous dissipation; c)
it possesses a term related to change in pressure, which some 
authors, e.g. \cite{Pons2005,Pons2007}, call the ``piston effect'',
and the resultant Boussinesq equations, the thermodynamic Boussinesq
equations.

\par

  In practice, the pressure effect can be accounted for by constructing 
evolution equations for the potential temperature or conservative temperature
respectively. 
  In order to show how an equation for $\Theta$ can be constructed, it is
useful to construct the potential temperature $\theta$ first. By definition,
$\theta$ is the temperature that a parcel with temperature $T$ would have
if lifted adiabatically to a reference level $Z=0$. Potential temperature is
thus defined as the implicit solution of the following equation:
\begin{equation}
       \eta_{ba} (T,S,Z) = \eta_{ba} (\theta,S,0) .
\end{equation}
Differentiating this expression yields:
\begin{equation}
      {\rm d}\eta_{ba} = \frac{\tilde{c}_p}{T} {\rm d}T 
    - \frac{\partial \tilde{\mu}}{\partial T} {\rm d}S  
     + \tilde{\alpha} g_0 {\rm d}Z = \frac{\tilde{c}_p^r}{\theta}{\rm d}\theta 
    - \frac{\partial \tilde{\mu}^r}{\partial S} {\rm d}S ,
\end{equation}
where we defined $\tilde{c}_p^r = c_p (\theta,S,0)$ and $\tilde{\mu}^r 
= \tilde{\mu}(\theta,S,0)$.
In the following, we simplify notations by defining $\tilde{c}_p^r=
c_p(\theta,0)$.
This expression shows that it is possible to rewrite the evolution
equation for entropy as follows:
\begin{equation}
      \frac{\tilde{c}_p^r}{\theta}\frac{D\theta}{Dt} = 
      \frac{\partial \tilde{\mu}}{\partial S} \frac{DS}{Dt}
      + \frac{D\eta_{ba}}{Dt}
\end{equation}
In absence of salinity, we can write:
\begin{equation}
     \frac{\tilde{c}_p^r}{\theta}\frac{D\theta}{Dt} = 
     \frac{\nabla \cdot ( \kappa \tilde{c}_p \nabla T )}{T} + 
     \frac{\varepsilon_K}{T} .
\end{equation}
This can be transformed in the following equation for $\theta$,
\begin{equation}
     \frac{D\theta}{Dt} = \nabla \cdot 
     \left ( \frac{\kappa \tilde{c}_p \theta \nabla T}{\tilde{c}_P^r T} 
 \right )
    + \dot{\theta}_{irr} ,
\end{equation}
where the nonconservative production/destruction of $\theta$ is
given by:
\begin{equation}
      \dot{\theta}_{irr} = - \kappa \tilde{c}_p \nabla T \cdot
      \nabla \left ( \frac{\theta}{\tilde{c}_p^r T} \right )
    + \frac{\theta \varepsilon_K}{\tilde{c}_p^r T} .
\end{equation}
In the literature, $\theta$ is often treated as a conservative 
variable, which consists in neglecting the nonconservative term
$\dot{\theta}_{irr}$. Alternatively, one may remark that the 
equation for $\theta$ can also be written as:
\begin{equation}
     \tilde{c}_{p}^r \frac{D\theta}{Dt} = \frac{D\tilde{h}_{\theta}}{Dt} , 
\end{equation}
where $\tilde{h}_{\theta}(\theta) = h_b(\eta,0)$ is the enthalpy 
a parcel would have if moved adiabatically from $Z$ to $Z=0$.
For this reason, $\tilde{h}_{\theta}$ was called ``potential enthalpy''
by \cite{McDougall2003}. This allows one to defined a new
temperature variable $\Theta$, also conserved for adiabatic
motions, such that: $d\tilde{h}_{\theta} = c_p^0 {\rm d}\Theta$, where
$c_p^0$ is an arbitrarily defined specific heat capacity 
representative of $c_p$ at $Z=0$. \cite{McDougall2003} 
discusses a possible choice for $c_p^0$ in the oceanic context.
As a result, it is possible to write the above equation as:
\begin{equation}
    \frac{D\Theta}{Dt} = \frac{1}{c_p^0} \frac{D\tilde{h}_{\theta}}{Dt}
   = \nabla \cdot 
  \left ( \frac{\kappa \tilde{c}_p \theta \nabla T}{c_p^0 T} \right )
   + \dot{\Theta}_{irr} ,
\end{equation}
where this time, the nonconservative production/destruction of 
$\Theta$ is given by:
\begin{equation}
  \dot{\Theta}_{irr} = - \frac{\kappa \tilde{c}_p}{c_p^0} \nabla T \cdot
  \nabla \left ( \frac{\theta}{T} \right ) + \frac{\theta \varepsilon_K}{T} .
\end{equation}
\cite{McDougall2003} provides convincing evidence that 
the nonconservative term $\dot{\Theta}_{irr}$ is significantly smaller
than $\dot{\theta}_{irr}$ for the present-day oceans. The relative 
smallness of $\dot{\Theta}_{irr}$ over $\dot{\theta}_{irr}$ does not appear
to be entirely generic though, and should therefore be checked on a case
by case basis, as it may depend on the particular fluid considered and on
the degree of turbulence present.
It is beyond the scope to simplify the equations further, and
to discuss when the evolution equation for $\Theta$ can be approximated
by the above simple diffusive model using $\kappa \nabla^2 T$.

\section{Thermodynamically and energetically 
consistent model for a Boussinesq
fluid with a linear equation of state}

\label{linear_eos}

  In this section, we return to the case of a Boussinesq fluid
with a linear equation of state, originally assumed to be governed
by Eqs. (\ref{velocity_model1}-\ref{eos_model1}), with the aim
of clarifying its thermodynamics, as well as to improve its formulation
to make it thermodynamically and energetically consistent.

\subsection{Thermodynamic properties of a Boussinesq fluid with a linear
equation of state}

  We first seek to construct the relevant forms of the specific Gibbs
function, enthalpy and internal energy, assuming a linear equation
of state $\rho = \rho_0 [1-\alpha (T-T_0)]$ and constant specific heat
capacity $c_{p0}$, so that the expression for the buoyancy is:
\begin{equation}
        b = \alpha g_0 (T-T_0) .
\end{equation}
 We need to decide on whether to interpret $T$
as the potential or in-situ temperature in the equation of state, as 
we saw above that any thermodynamically consistent formulation 
must distinguish between the two kinds of temperature. For 
completeness, the two cases are considered. We first discuss the
case where $T$ is the in-situ temperature, deferring the discussion
of an equation of state linear in potential temperature to 
subsection \ref{theta_case}. This choice amounts to assume that
the isothermal compressibility, rather than the adiabatic
compressibility, vanishes. From a thermodynamic viewpoint, this is
arguably an awkward assumption, because Eq.
(\ref{compressibilities}) suggest that this endows the fluid with a 
negative speed of sound, but since sound waves are filtered out,
it is unclear whether this is a serious impediment.

\par

 The interpretation of $T$ as the in-situ temperature motivates looking
for a description of the thermodynamic properties in terms of the 
specific Gibbs function $g_{ba}$, since the natural variables of the
latter are $T$ and $Z$. As seen previously, the total differential of
$g_{ba}$ is ${\rm d}g_{ba} = 
-\eta_{ba}{\rm d}T - b{\rm d}Z$, which implies:
\begin{equation}
    \frac{\partial g_{ba}}{\partial T} = -\eta_{ba}, \qquad
    \frac{\partial g_{ba}}{\partial Z} = - b .
   \label{gba_derivatives}
\end{equation}
The assumption of constant heat capacity provides the additional
differential relation:
\begin{equation}
      c_{p0} = -T \frac{\partial^2 g_{ba}}{\partial T^2} .
      \label{cpo_gibbs}
\end{equation}
The system of partial differential equations Eqs. (\ref{gba_derivatives})
and (\ref{cpo_gibbs}) is sufficient to completely specify $g_{ba}$, 
whose integration yields
\begin{equation}
      g_{ba} = - c_{p0} \left [ T \ln{(T/T_0)} - T \right ]
    - \alpha g_0 (Z-Z_0) (T-T_0) ,
\end{equation}
up to some arbitrary constant of integration. From 
Eq. (\ref{gba_derivatives}), the following expression for
the specific entropy is obtained:
\begin{equation}
      \eta_{ba} = -\frac{\partial g_{ba}}{\partial T} = 
        c_{p0} \ln{\left ( \frac{T}{T_0} \right )} + \alpha g_0 (Z-Z_0).
      \label{boussinesq_entropy}
\end{equation}
This makes it possible to derive an exact expression for the potential
temperature $\theta$, which is the implicit solution of the equation
$\eta_{ba}(T,Z) = \eta_{ba}(\theta,0)$, i.e., 
$c_{p0}\ln{(T/T0)} + \alpha g_0 (Z-Z_0) = c_{p0} \ln{(\theta/T_0)}
-\alpha g_0 Z_0$,
yielding:
\begin{equation}
   \theta = T \exp{\left \{ \frac{\alpha g Z}{c_{p0}} \right \}}
\end{equation}
This makes it clear that in order for a model to be thermodynamically
consistent, potential temperature is always different from the 
in-situ temperature. Now, we previously established that the 
expressions for internal energy and enthalpy were given by:
$e_{ba} = g_{ba}+ b(Z-Z_0) + T \eta_{ba}$ and 
$h_{ba} = e_{ba} - b (Z-Z_0) = g_{ba} + T \eta_{ba}$, yielding 
therefore the following expressions:
\begin{equation}
      h_{ba} = c_{p0} T + \alpha g_0 T_0 (Z-Z_0) ,
\end{equation}
\begin{equation}
      e_{ba} = c_{p0} \left [ 1 + \frac{\alpha g_0(Z-Z_0)}{c_{p0}}
      \right ] T .
\end{equation}
These relations can be written in terms of other variables by
using the expression for $\theta$, as well as $b$. For 
instance, the enthalpy can be written in terms of the 
potential temperature and $Z$ as follows:
\begin{equation}
     h_{ba} = c_{p0} \theta \exp{\left \{ - \frac{\alpha g_0 Z}
     {c_{p0}} \right \}} + \alpha g_0 T_0 (Z-Z_0) .
\end{equation}
\begin{equation}
      e_{ba} = c_{p0} \left [ 1 - \frac{\alpha g_0 Z_0}{c_{p0}} +
     \ln{\left \{ 
      \frac{\alpha g_0 \theta}{b + \alpha g_0 T_0} \right \}}
      \right ] \left ( T_0 + \frac{b}{\alpha g_0} \right )
\end{equation}

\subsection{Improvement of the model energetic and thermodynamic consistency}

 Having clarified the nature of the thermodynamics and 
thermodynamic potentials for a Boussinesq fluid with a 
linear equation, we now turn to the issue of writing down
explicitly how to transform the initial formulation into
an energetically and thermodynamically consistent one.
The main modification introduced is related to the heat
equation. Moreover, there is now a distinction between
in-situ temperature and potential temperature. To summarise,
the whole model is therefore given as follows:
\begin{equation}
   \frac{D{\bf v}}{Dt} + \frac{1}{\rho_0}
   \nabla \delta P = b {\bf k} + \nu \nabla^2 {\bf v}
\end{equation}
\begin{equation}
     b = \alpha g_0 (T-T_0)
\end{equation}
\begin{equation}
    \nabla \cdot {\bf v} = 0
\end{equation}
\begin{equation}
    \frac{D\theta}{Dt} = \frac{\theta}{T} \kappa \nabla^2 T 
  + \frac{\theta \varepsilon_K}{c_{p0} T}
  \label{theta_equation}
\end{equation}
\begin{equation}
     T = \theta \exp{\left \{ -\frac{\alpha g_0 Z}{c_{p0}} 
     \right \}}
     \label{T_theta}
\end{equation}
Note that we can rewrite the temperature equation as:
\begin{equation}
     \frac{D\theta}{Dt} = \nabla \cdot \left ( 
     \frac{\kappa \theta}{T} \nabla T \right ) + \dot{\theta}_{irr}
     + \frac{\theta \varepsilon_K}{c_{p0}T}
\end{equation}
where 
\begin{equation}
     \dot{\theta}_{irr} = -\kappa \nabla T \cdot 
     \nabla \left ( \frac{\theta}{T} \right ) 
     = -\frac{\kappa \alpha g_0}{c_{p0}} \frac{\theta}{T} 
     \frac{\partial T}{\partial z} 
     = - \frac{\kappa \alpha g_0}{c_{p0}} \frac{\partial \theta}{\partial z}
     + \frac{\kappa \alpha^2 g_0^2}{c_{p0}^2} \theta
\end{equation}
Note that it is also possible to write the flux entirely in terms of
$\theta$:
\begin{equation}
     \frac{\kappa \theta}{T} \nabla T = 
     \kappa \nabla \theta - \frac{\kappa \alpha g_0 \theta}{c_{p0}}
 {\bf k}
\end{equation}
which was computed assuming $Z(z)=z$.
Interestingly, these equations introduce
the following length scale: $L = c_{p0}/(\alpha g_0)$. For
typical values, $c_{p0}=4.10^3 {\rm J.kg^{-1}.K^{-1}}$, 
$\alpha = 10^{-4}\,{\rm K}^{-1}$ and $g_0 = 10 \,
{\rm m.s^{-1}}$, which yields: $L = 4.10^6\,{\rm m}$.
As $L$ is huge compared to the typical length scales at which
molecular diffusion is important, it follows that the nonconservative
term is probably negligible.

\subsection{Equation of state linear in potential temperature}
\label{theta_case}

 This section revisits the above results by assuming that the equation
of state is linear in $\theta$ rather than in $T$, and hence that the
adiabatic compressibility vanishes, as is generally implicitly assumed
in traditional low Mach number asymptotics. In this case, the buoyancy
becomes:
\begin{equation}
      b = \alpha_{\theta} g_0 (\theta-\theta_0) ,
\end{equation}
with $\alpha_{\theta}$ the isentropic thermal expansion,
which \cite{Tailleux2010} shows (his Eq. A.11) is related to the classical
thermal expansion by $\alpha_{\theta} = \alpha T/\theta$ for
a constant $c_{p0}$. The use of $\theta$ as a dependent variable
makes it natural to seek a description of the thermodynamic properties
of the fluid from the specific enthalpy, since $\theta$ is closely related
to the specific entropy by ${\rm d}\eta_{ba} = c_{p0} {\rm d}\theta/\theta$,
as seen previously. As a result, the total differential of specific enthalpy
can be written as:
\begin{equation}
   dh_{ba} = T {\rm d}\eta_{ba} - b {\rm d}Z =
    \frac{T c_{p0}}{\theta} {\rm d}\theta - b {\rm d}Z ,
\end{equation}
which yields the following system of two partial differential
equations:
\begin{equation}
       \frac{\partial h_{ba}}{\partial \theta} = \frac{c_{p0}T}{\theta},
       \qquad \frac{\partial h_{ba}}{\partial Z}  = -b = - \alpha_{\theta}
       g_0 (\theta-\theta_0) .
       \label{hba_pde}
\end{equation}
The Maxwell relationship for such a system imposes the following
integrability constraint:
\begin{equation}
       \frac{\partial}{\partial Z}\left ( \frac{c_{p0}T}{\theta} \right )
     = - \frac{\partial b}{\partial \theta} = -\alpha_{\theta} g_0 ,
\end{equation}
which can be integrated to yield the following expression 
between $T$ and $\theta$:
\begin{equation}
       \frac{T}{\theta} = 1 -\frac{\alpha_{\theta} g_0 Z}{c_{p0}},
       \label{T_theta_2}
\end{equation}
using the fact that by definition, $T=\theta$ at $Z=0$.
By using Eq. (\ref{T_theta_2}), it becomes straightforward to 
integrate the partial differential relations Eqs. (\ref{hba_pde})
to eventually obtain
\begin{equation}
       h_{ba} = 
   c_{p0}\left [ 1 - \frac{\alpha_{\theta} g_0 Z}
  {c_{p0}} \right ] (\theta - \theta_0) = 
  \left ( \frac{\theta-\theta_0}{\theta} \right ) c_{p0} T,
\end{equation}
up to an arbitrary constant of integration. The expression for the
internal energy is therefore given by:
\begin{equation}
     e_{ba} = h_{ba} + b (Z-Z_0) = [c_{p0}-\alpha_{\theta}g_0 Z_0]
    (\theta-\theta_0) ,
\end{equation}
which implies that in the present model, $\theta$ is in fact a
proxy for the specific internal energy. It is useful to note here that
because $c_{p0}$ is constant, the concepts of potential 
temperature $\theta$ and conservative temperature $\Theta$
coincide. 

\par

  The evolution equation for temperature is unchanged, while the
expression linking $T$ and $\theta$ is now given by Eq. (\ref{T_theta_2}),
so that in summary, the thermodynamics is now described by the system:
\begin{equation}
     \frac{D\theta}{Dt} = \frac{\theta}{T} \kappa \nabla^2 T
  + \frac{\theta \varepsilon_K}{c_{p0}T} ,
\end{equation}
\begin{equation}
      T = \theta \left [ 1 - 
   \frac{\alpha_{\theta} g_0 Z}{c_{p0}} \right ] .
\end{equation}
As shown in this paper, the diffusive term can be written as the
sum of the divergence of a diffusive flux, plus a nonconservative
production/destruction term, which are explicitly given by:
\begin{equation}
     \frac{D\theta}{Dt} = \nabla \cdot \left ( \frac{\kappa \theta}{T} \nabla T \right )
     + \dot{\theta}_{irr} + \frac{\theta \varepsilon_K}{c_{p0}T}
\end{equation}
where
\begin{equation}     
       \dot{\theta}_{irr} = 
       - \kappa \nabla T \cdot \nabla \left ( \frac{\theta}{T} \right ) 
  = 
     - \frac{\kappa \alpha_{\theta} g_0/c_{p0}}{
     \left [ 1 - \alpha_{\theta} g_0 Z/c_{p0} \right ]^2} 
     \frac{\partial T}{\partial z} = 
   - \frac{\kappa \alpha_{\theta} g_0}{c_{p0}}
      \left ( \frac{\theta}{T}\right )^2 \frac{\partial T}{\partial z}
\end{equation}
The nonconservative terms have been shown by \cite{McDougall2003}
to be very small compared to the diffusive term, so that can be neglected
in practice. 

\section{Thermodynamically and energetically consistent Boussinesq
primitive equation ocean models}

\label{nonlinear_eos}

  In this section, we return to the Boussinesq primitive equations
Eqs. (\ref{velocity_ogcm}-\ref{eos_ogcm}) that form the basis for
most current numerical ocean general circulation models (save
for the form of the turbulent parameterisations), with the aim of 
showing how to modify them in order to make them energetically
and thermodynamically consistent.

\subsection{Improved formulation}

  It should be clear by now from the above results that two main
modifications are needed to make the Boussinesq primitive equations
(\ref{velocity_ogcm}-\ref{eos_ogcm}) energetically and thermodynamically
consistent, which are: 1) modification of the definition of buoyancy, 2)
addition of the nonconservative terms in the thermodynamic equation
Eq. (\ref{temperature_ogcm}). The resulting set of modified equations
is as follows:
\begin{equation}
    \frac{D{\bf u}}{Dt} + f {\bf k} \times {\bf u} + 
    \frac{1}{\rho_0} \nabla_h \delta P = 
    A_H \nabla_h^2 {\bf u} + A_V \frac{\partial^2 {\bf u}}{\partial z^2} ,
\end{equation}
\begin{equation}
    \frac{1}{\rho_0} \frac{\partial \delta P}{\partial z} = 
    - g_0 \left ( \frac{\tilde{\rho}-\rho_0}{\tilde{\rho}} \right )
    \frac{\partial Z}{\partial z} = b \frac{\partial Z}{\partial z} ,
\end{equation}
\begin{equation}
     \nabla_h \cdot {\bf u} + \frac{\partial w}{\partial z} = 0 ,
\end{equation}
\begin{equation}
    \frac{D\Theta}{Dt} = K_H \nabla_h^2 \Theta + 
    K_v \frac{\partial^2 \Theta}{\partial z^2} +
    \dot{\Theta}_{irr} ,
\end{equation}
\begin{equation}
     \frac{DS}{Dt} = K_H \nabla_h^2 S + K_v \frac{\partial^2 S}{\partial z^2} ,
\end{equation}
\begin{equation}
       \tilde{\rho} = \rho(S,\Theta,P_0(Z)) ,
\end{equation}
where $\Theta$ is now assumed to be the conservative temperature, 
on the basis of \cite{McDougall2003}'s results suggesting that 
the nonconservative term $\dot{\Theta}_{irr}$ is significantly smaller
for $\Theta$ than for the potential temperature $\theta$, resulting in a
smaller error in the overall energy budget when this term is neglected.
For completeness, we need to provide an explicit expression for 
$\dot{\Theta}_{irr}$. To that end, it is first necessary to understand 
how to evaluate a number of thermodynamic properties for 
Boussinesq seawater, which is discussed next.

\subsection{Thermodynamics of Boussinesq seawater}

  In contrast to the idealised Boussinesq model of the previous section,
the aim of numerical ocean modelling is to achieve realistic simulations,
and therefore to use an equation of state that is as realistic as possible.
As mentioned earlier, the thermodynamic properties of seawater are 
currently encapsulated in a Gibbs function, e.g., 
\cite{Feistel2003,teos10}, with natural variables in-situ temperature,
salinity, and pressure. It is therefore natural to seek a determination of
Boussinesq thermodynamics from the published Gibbs function.
The Boussinesq Gibbs function was shown above to be given by:
\begin{equation}
      g_{ba}(T,S,Z) = g(T,S,P_0(Z)) + g_0 (Z-Z_0) = \tilde{g} + 
     g_0(Z-Z_0) .
\end{equation}
As is well known, the specific entropy $\eta_{ba}$, the relative 
chemical potential $\mu_{ba}$, the buoyancy $b$ (and specific
volume), and the specific heat capacity $c_{p,ba}$ can all be 
obtained from the first and second partial derivatives of $g_{ba}$ 
as follows:
\begin{equation}
     \eta_{ba} (T,S,Z) = \tilde{\eta} = 
       -\frac{\partial \tilde{g}}{\partial T} ,
\end{equation}
\begin{equation}
      \mu_{ba}(T,S,Z) = \tilde{\mu} = 
    \frac{\partial \tilde{g}}{\partial S} ,
\end{equation}
\begin{equation}
       b = - \frac{\partial g_{ba}}{\partial Z} = 
       -\frac{\partial \tilde{g}}{\partial Z} - g_0 
         = g_0(\rho_0 \tilde{\upsilon}-1) ,
\end{equation}
\begin{equation}
        c_{pa} = \tilde{c} = - T \frac{\partial^2 \tilde{g}}{\partial T^2},
\end{equation}
where the formula for $b$ was arrived at by using the result that
$\partial g/\partial P = \upsilon(S,T,P)$. As previously, tilded quantities
refer to quantities estimated for the hydrostatic reference pressure
$P_0(Z)$ rather than the full pressure, so that 
$\tilde{\upsilon}=\upsilon(T,S,P_0(Z))$ for instance.

\par

In order to arrive at an expression for the conservative temperature,
we need the expressions for the specific enthalpy $h_{ba}$ and potential 
enthalpy $h_{\theta}$, which are respectively given by:
\begin{equation}
    h_{ba}(T,S,Z) = g_{ba}(T,S,Z) + T \tilde{\eta} = \tilde{g} + 
    g_0 (Z-Z_0) + T \tilde{\eta} ,
\end{equation}
\begin{equation}
       h_{\theta} (T,S,Z) = h_{ba}(\theta,S,0) = h_r (\theta,S) ,
      \label{hba_theta}
\end{equation}
where the potential temperature $\theta$ is now the implicit solution of
$\eta_{ba}(T,S,Z) = \eta_{ba} (\theta,S,0)$. Using the above expressions
for the specific entropy and heat capacity, as well as the Maxwell 
relationships for the Gibbs function, yields the following expression for
the total differential of specific entropy:
\begin{equation}
    {\rm d}\eta_{ba} = {\rm d}\tilde{\eta} = 
     \frac{\tilde{c}_p {\rm d}T}{T} - \frac{\partial \tilde{\mu}}{\partial T}
     {\rm d}S + \frac{\partial b}{\partial T} {\rm d}Z ,
    \label{entropy_differential}
\end{equation}
which in turn yields the following expression for
 the total differential of specific enthalpy:
$$
   {\rm d}h_{ba} = T {\rm d}\tilde{\eta} + \tilde{\mu}{\rm d}S
    - b {\rm d}Z
      = T \underbrace{\left [ \tilde{c}_p \frac{{\rm d}T}{T} - 
    \frac{\partial \tilde{\mu}}{\partial T} {\rm d}S 
    + \frac{\partial b}{\partial T} {\rm d}Z \right ]}_{{\rm d}\tilde{\eta}}
     + \tilde{\mu} {\rm d}S - b {\rm d}Z 
$$
\begin{equation}
    = \tilde{c}_p {\rm d}T + \left ( \tilde{\mu} - T 
    \frac{\partial \tilde{\mu}}{\partial T} \right ) {\rm d}S  
   + \left ( g_0 \tilde{\alpha} T - b \right ){\rm d}Z ,
\end{equation}
where $\partial b/\partial T = \rho_0 g_0 \partial \tilde{\upsilon}/
\partial T = g_0 \tilde{\alpha}$, by defining the thermal expansion
coefficient by $\tilde{\alpha} = \rho_0 \partial \tilde{\upsilon}/
\partial T$. Similarly, the haline contraction coefficient should
be defined by $\tilde{\beta} = -\rho_0 \partial \tilde{\upsilon}/
\partial S$. By evaluating the above expression at $Z=0$, 
we obtain the following expression for the total differential of potential 
enthalpy:
\begin{equation}
     {\rm d} h_{\theta} = \tilde{c}_p^r {\rm d} \theta
   + \left ( \tilde{\mu}_r - \theta \frac{\partial \tilde{\mu}_r}
   {\partial \theta} \right ) {\rm d}S = c_p^0 {\rm d}\Theta ,
\end{equation}
where $\tilde{c}_p^r = \tilde{c}_p(\theta,S,0)$ and
$\tilde{\mu}_r = \tilde{\mu}(\theta,S,0)$. As discussed by 
\cite{McDougall2003}, the conservative temperature is defined such
that ${\rm d}\Theta = {\rm d}h_{\theta}/c_p^0$, for some constant
representative value of the specific heat capacity $c_p^0$.
The latter expression allows one to express
${\rm d}\theta$ in terms of ${\rm d}\Theta$ as follows:
\begin{equation}
      {\rm d}\theta = \frac{c_p^0}{\tilde{c}_p^r} {\rm d}\Theta
      - \frac{1}{\tilde{c}_p^r} \left ( 
      \tilde{\mu}_r - \theta \frac{\partial \tilde{\mu}}{\partial \theta}
     \right ) {\rm d}S .
    \label{potential_to_conservative}
\end{equation}
Now, by evaluating the differential of entropy Eq. 
(\ref{entropy_differential}) at $Z=0$, as well as by using the 
implicit definition of potential temperature and Eq. 
(\ref{potential_to_conservative}), it is possible to write down the
differential of entropy equivalently in terms of $\theta$ or $\Theta$
as follows:
\begin{equation}
     {\rm d}\tilde{\eta} = \frac{\tilde{c}_p^r}{\theta}{\rm d}\theta
     - \frac{\partial \tilde{\mu}_r}{\partial \theta} {\rm d}S
   =  \frac{c_p^0}{\theta}{\rm d}\Theta 
      - \frac{\tilde{\mu}_r}{\theta} {\rm d}S ,
\end{equation}
which in turn allows one to write the total differential of specific
enthalpy in terms of the conservative temperature as follows:
\begin{equation}
     {\rm d}h_{ba} = T {\rm d}\tilde{\eta} + \tilde{\mu}{\rm d}S - b{\rm d}Z
    = \frac{Tc_p^0}{\theta} {\rm d}\Theta 
  + \left ( \tilde{\mu} - \frac{T\tilde{\mu}_r}{\theta} \right ) {\rm d}S
  - b {\rm d}Z .
   \label{first_law_Theta}
\end{equation}
This result is used next to provide an explicit expression for the
nonconservative production of conservative temperature.

\subsection{Determination of the nonconservative term $\dot{\Theta}_{irr}$}

 As discussed in \cite{Tailleux2010}, the determination of
$\dot{\Theta}_{irr}$ follows from the constraint of total energy 
conservation. In order to see this, let us first derive the 
evolution equation for the kinetic energy by multiplying the
momentum equations by ${\bf v}$. After some manipulation, this
can be put in the form:
\begin{equation}
    \rho_0 \frac{D}{Dt} \frac{{\bf u}^2}{2} 
    + \nabla \cdot (\delta P {\bf v} -\rho_0 {\bf F}_{ke} ) = 
   \rho_0 b \frac{DZ}{Dt} - \rho_0 \varepsilon_K ,
   \label{ke_ogcm}
\end{equation}
where the diffusive flux of kinetic energy and viscous dissipation
are given by:
\begin{equation}
       {\bf F}_{ke} = A_H \nabla_h \frac{{\bf u}^2}{2} +
       A_v \frac{\partial}{\partial z} \frac{{\bf u}^2}{2} {\bf k} ,
\end{equation}
\begin{equation}
      \varepsilon_K =  A_h [ \| \nabla_h u\|^2 + \| \nabla_h v\|^2 ] 
      + A_v \left ( \frac{\partial {\bf u}}{\partial z} \right )^2 .
\end{equation}
Now, by inserting the evolution equations for $\Theta$ and $S$ into
Eq. (\ref{first_law_Theta}), it is possible to rewrite the evolution
equation for the specific enthalpy as follows:
$$
   \rho_0 \frac{Dh_{ba}}{Dt} = 
  \frac{T c_p^0}{\theta} \rho_0 \frac{D\Theta}{Dt}
 + \rho_0 \left ( \tilde{\mu} - \frac{T\tilde{\mu}_r}{\theta} \right )
   \frac{DS}{Dt} - \rho_0 b \frac{DZ}{Dt} 
$$
\begin{equation}
   = \nabla \cdot ( \rho_0 {\bf F}_h ) - \rho_0 \dot{h}_{irr}
     + \frac{T c_p^0}{\theta} \rho_0 \dot{\Theta}_{irr} 
    - \rho_0 b \frac{DZ}{Dt} ,
   \label{enthalpy_ogcm}
\end{equation}
where the diffusive flux ${\bf F}_h$ and the irreversible 
nonconservative term $\dot{h}_{irr}$ are given by:
\begin{equation}
      {\bf F}_h = \frac{Tc_p^0}{\theta} 
    \left [ K_H \nabla_h \Theta + K_v \frac{\partial \Theta}{\partial z}
     {\bf k} \right ] 
     + \left ( \tilde{\mu}_r - \frac{T \tilde{\mu}_r}{\theta} 
     \right ) \left [ K_H \nabla_h S + K_v \frac{\partial S}{\partial z}
    {\bf k} \right ] ,
\end{equation}
\begin{equation}
     \dot{h}_{irr} =  K_H 
 \left [ \nabla_h S \cdot \nabla \tilde{\mu}^{\ast \ast} + 
         \nabla_h \Theta \cdot \nabla_h \left ( \frac{T c_p^0}{\theta}
      \right ) \right ] + K_v
  \left [ \frac{\partial S}{\partial z} \frac{\partial 
   \tilde{\mu}^{\ast \ast}}
   {\partial z} + \frac{\partial \Theta}{\partial z} 
  \frac{\partial}{\partial z} \left ( \frac{T c_p^0}{\theta} \right )
   \right ], 
\end{equation}
where $\tilde{\mu}^{\ast \ast} = \tilde{\mu} 
- T \tilde{\mu}_r/\theta$. Finally, an evolution equation for the total
energy is obtained by Eqs. (\ref{ke_ogcm}) 
and (\ref{enthalpy_ogcm}),
\begin{equation}
   \rho_0 \frac{D}{Dt} \left ( \frac{{\bf v}^2}{2} + h_{ba}
  \right ) + \nabla \cdot ( \delta P {\bf v} 
  + \rho_0 {\bf F}_h - \rho_0 {\bf F}_{ke} ) = 
  \rho_0 \left ( \dot{h}_{irr} + \frac{Tc_p^0}{\theta}\dot{\Theta}_{irr}
    - \rho_0 \varepsilon_K \right ),
\end{equation}
which shows that the total energy 
${\bf u}^2/2+h_{ba}$ is a conservative quantity only if the
right-hand side of the later equation vanishes, which yields:
\begin{equation}
       \dot{\Theta}_{irr} =  \frac{\theta}{c_p^0 T} 
   \left [ \varepsilon_K + \dot{h}_{irr} \right ] .
\end{equation}
This completes the full specification of the modified Boussinesq
primitive equations.

\subsection{Errors in the energy budget of current OGCMs}

  The above results imply that there are at least two
main sources of error in the global energy budget of current OGCMs, 
resulting respectively
from their use of the buoyancy 
$b_{ogcm}=-g_0(\tilde{\rho}-\rho_0)/\rho_0$ instead
of $b=-g_0(\tilde{\rho}-\rho_0)/\tilde{\rho}$, as well as from 
neglecting the irreversible production term $\dot{\Theta}_{irr}$ in the
temperature equation (see \cite{Tailleux2010} for relevant expressions
when potential temperature is used). Thus, by replacing the buoyancy
$b$ by $b_{OGCM}$ in the kinetic energy equation, as well as by 
neglecting $\dot{\Theta}_{irr}$ in the enthalpy equation, implies 
for the evolution equation of total energy in current OGCMs to be
of the form:
\begin{equation}
     \rho_0 \frac{D}{Dt} \left ( \frac{{\bf v}^2}{2} + h_{ba} \right )
 + \nabla \cdot \left ( \delta P {\bf v} + \rho_0 {\bf F}_h - \rho_0 
 {\bf F}_{ke} \right ) = \rho_0 \dot{E}_{irr} ,
\end{equation}
where the right-hand side $\rho_0 \dot{E}_{irr}$ corresponds to the
spurious production/destruction of total energy resulting from the
inconsistent treatment of the buoyancy and temperature equation.
From the above results, its explicit expression is given by:
\begin{equation}
     \rho_0 \dot{E}_{irr} = -\rho_0 \left ( \dot{h}_{irr} +
     \varepsilon_K \right ) + \rho_0 (b_{ogcm}-b) w
     \approx -\rho_0 \left ( \dot{h}_{irr} + \varepsilon_K \right )
    - \left ( \frac{\tilde{\rho}-\rho_0}{\rho_0} \right )^2 
   \rho_0 w ,
\end{equation}
since we have:
\begin{equation}
     b_{ogcm} - b = -g_0 \left ( \frac{\tilde{\rho}-\rho_0}{\rho_0} \right )
     + g_0 \left ( \frac{\tilde{\rho}-\rho_0}{\tilde{\rho}} \right )
    = g_0 \left ( 2 - \frac{\tilde{\rho}}{\rho_0} - \frac{\rho_0}{\tilde{\rho}}
   \right ) \approx - \left ( \frac{\tilde{\rho}-\rho_0}{\rho_0} \right )^2 . 
\end{equation}
Note here that the final form for the error term $\dot{E}_{irr}$ 
differs significantly from that discussed in \cite{Tailleux2010}, 
because in the latter study, $\dot{E}_{irr}$ was estimated using
the non-approximated thermodynamic potentials, resulting in additional
spurious sinks/sources of energy. Interestingly, it is important to 
note that because the term $(b_{ogcm}-b)$ scales as the square of
the small Boussinesq parameter $(\rho-\rho_0)/\rho_0$, it follows
that the energetically and thermodynamically consistent Boussinesq
formulation has the same order of accuracy as the non-consistent
formulation. From that viewpoint, it can be said that of all the
possible Boussinesq approximations that can be constructed with
the same formal order of accuracy, one can be found that is energetically
and thermodynamically consistent.

\par

  As discussed in \cite{Tailleux2010}, errors in the global energy
are likely to be of consequence only if they are associated with 
spurious forces in the momentum equations. 
Indeed, \cite{Tailleux2010} estimated
that the same error $O(10^{-6}\,{\rm W.m^{-3}})$ in the overall energy
budget corresponds only to a spurious heat source/sink in the 
temperature equation that is $O(7.5 K/({\rm million\,\,years}))$,
which is arguably utterly small, whereas it may correspond to spurious
positive or negative acceleration in the momentum equations 
$O(3 {\rm m.s^{-1}}/{\rm year})$ in the momentum equations, which is
in contrast a significant number. On this basis, it appears important
to correct the definition of the buoyancy in current OGCMs, but less
vital to retain the nonconservative term $\dot{\Theta}_{irr}$. 
From a practical computational viewpoint, modifying the definition of
buoyancy in current OGCM implementations should be rather 
straightforward, as it amounts to change the integrand in the 
particular sub-routine estimating the hydrostatic pressure from the
knowledge of density. By contrast, evaluating the nonconservative
term $\dot{\Theta}_{irr}$ at each grid point at all time steps would
be a major undertaking, since it requires the evaluation of many
thermodynamic quantities such as
 $\tilde{\mu}$, $\tilde{\mu}_r$, $\tilde{c}_p^r$, $\theta$ and $T$.
This would be computationally prohibitive, given that the evaluation 
of density alone already amounts for a significant fraction of the total
CPU time owing to its strongly nonlinear character.

\section{Summary and discussion}

  In this paper, we sought to clarify the nature of the conversions between 
mechanical energy and internal energy
supported by the Boussinesq and anelastic approximations,
in the general case of a binary fluid with an arbitrary nonlinear equation
of state. A key result was to show that the energetics of such approximations
possesses a term that can be identified as playing the role of the classical
compressible work of expansion/contraction, which manifests itself as apparent
changes in gravitational potential energy due changes in density (and hence
of mass if the volume is exactly or approximately conserved, as is the case
for such approximations). In contrast with a fully compressible fluid, however,
the conversion between mechanical energy and internal energy is between
GPE and IE rather than between KE and IE. By regarding this term as the
Boussinesq/anelastic approximation to the compressible work
$\delta W = -P {\rm d}\upsilon$, it is possible to construct the relevant 
approximation to the ``heat'' $\delta Q$ in a consistent way by ensuring
satisfaction of Maxwell relationships, which upon integration eventually leads
to the construction of consistent expressions for the specific internal energy,
as well as of the whole range of known thermodynamic potentials, illustrating
\cite{Bannon1996}'s statement that the ``thermodynamics is slaved to the 
dynamics''. 

\par

   The existence of well-defined thermodynamic potentials from which
to derive physically consistent expressions for the first law of 
thermodynamics appears to be sufficient to endow the Boussinesq and
anelastic approximations considered in this paper with fully consistent
energetics and thermodynamics, even when 
diabatic effects and a fully nonlinear equation of state for a binary
fluid are retained. In the energetically consistent form of the Boussinesq
and anelastic approximations, the
sum of kinetic energy and enthalpy is a conservative
quantity, and hence the natural total energy for the system. As a 
consequence, the gravitational potential energy can be regarded as the
difference between enthalpy and internal energy, and hence as a purely
thermodynamic property of the fluid. These ideas were illustrated by 
showing how two widely used but energetically inconsistent Boussinesq 
models could be modified to make them fully energetically and 
thermodynamically consistent. Interestingly,
we find that the modifications required to ensure
energetic and thermodynamic consistency do not alter the formal order
of accuracy of the approximations. 
In other words, in the space of all Boussinesq and anelastic approximations 
of a given order of accuracy, one exists that is fully consistent energetically
and thermodynamically. It was also showed how to construct explicitly the
full range of thermodynamic potentials for Boussinesq and anelastic models,
either by integrating Maxwell relationships in the context of idealised
models with idealised equations of state and heat capacities, or by 
approximating the exact thermodynamic potentials when those are known.
A direct application of our results is to suggest that current
numerical ocean circulation models possess a potentially significant 
source of error in their momentum equations owing to their use of an 
incorrect definition of buoyancy, which could in principle be simply
corrected by using the correct definition. Interestingly, the improved
definition of buoyancy is one that was recently proposed earlier by
 \cite{Pauluis2008} and \cite{Young2010}, but it does not appear to have
been realized until now that such a modification was needed to 
improve the energetic consistency of current OGCMs.

\par

  An important implication of our results is to support the earlier 
suggestion by \cite{Tailleux2009} that the Boussinesq (and hence 
anelastic) approximations can support large conversions between mechanical
energy and internal energy, and therefore compressibility effects 
significantly larger than previously assumed. Specifically, the point
made in \cite{Tailleux2009} is that in the context of turbulent stratified
mixing, the apparent irreversible conversion
of available GPE into background GPE should not be interpreted as a
mechanical to mechanical energy conversion, as proposed by \cite{Winters1995},
but as an irreversible conversion of AGPE into IE. Note, indeed, that
on the one hand, \cite{Winters1995} interpret the following energy 
conversion term:
\begin{equation}
      W_{r,laminar} = \int_{V} g_0 z \kappa \nabla^2 \rho {\rm d}V =
     \kappa g_0 \left [ \langle \rho \rangle_{bottom} 
     - \langle \rho \rangle_{top} \right ] ,
     \label{laminar_conversion}
\end{equation}
as a (laminar) conversion of internal energy into background gravitational
potential energy, but on the other hand, interpret the following energy
conversion term:
\begin{equation}
     W_{r,turbulent} =
   \int_{V} g z_r \kappa \nabla^2 \rho {\rm d} V - W_{r,laminar}
= - \int_{V} \frac{g \kappa \| \nabla \rho_r \|^2}
   {\partial \rho_r/\partial z_r}
   {\rm d}V - W_{r,laminar},
      \label{turbulent_conversion}
\end{equation}
as the irreversible conversion of AGPE into background $GPE_r$,
where $z_r = z_r({\bf x},t)$ is the parcel's position in Lorenz's reference
state, and hence as a mechanical to mechanical energy conversion, on the
grounds that $W_{r,turbulent}$ shows up in the evolution equations for
$AGPE$ and $GPE_r$ with opposite signs. Yet, both $W_{r,laminar}$ and
$W_{r,turbulent}$ are seen to involve terms of the form 
$g z \kappa \nabla^2 \rho$, which the present paper argues is the one
playing the role of the compressible work of expansion/contraction in
the Boussinesq/anelastic approximations. For this reason, 
\cite{Tailleux2009} argue that $W_{r,turbulent}$ actually refers to 
two different types of energy conversions in the $AGPE$ and $GPE_r$ 
evolution equations, for which two different notations should be used.
\cite{Tailleux2009} used the notation $D(APE)$ to refer to the dissipation
of AGPE into $IE$.

\par

   If one accepts the idea that $D(APE)$ and $W_{r,turbulent}$ actually
represent two large conversions between $AGPE$ and $IE$, and between
$IE$ and $GPE_r$ respectively, then the question arises as to how these
conversions are actually achieved in reality since there is no direct 
conversion between IE and GPE in the classical description of the 
energetics of the fully compressible Navier-Stokes equations? 
With regard to $W_{r,mixing}$, the theory of the hydrostatic adjustment,
e.g., \cite{Bannon1995} can be tentatively invoked to speculate on some
of the physical mechanisms and processes involved. Physically, localised
heating/cooling anomalies due to molecular diffusion must cause localised
pressure anomalies, which will propagate as acoustic waves whose energy
can be converted into kinetic energy via the compressible work of 
expansion/contraction and then ultimately into gravitational potential 
energy via the buoyancy flux $\rho g w$. If this is what indeed happens,
then it is interesting to note that the Boussinesq/anelastic approximations
implicitly assumes that of all the internal energy lost to mechanical 
energy, all of it goes toward increasing the gravitational potential energy,
whereas in reality, it seems plausible that some of it could go toward
increasing the turbulent kinetic energy, thereby acting as a positive 
feedback on turbulent mixing. The latter hypothesis warrants further
research, as acoustic waves are known to be capable of generating 
mean flows via acoustic streaming for instance, e.g., 
see \cite{Lighthill1978a,Vanneste2001}. What are the physical mechanisms
involved in the opposite conversion whereby $AGPE$ is ultimately 
dissipated into $IE$ by molecular diffusion is less clear, because
physically AGPE can only be converted reversibly into KE by construction.
As a result, the only way to dissipate AGPE into IE by molecular diffusive
processes seems to require converting AGPE into KE reversibly, then KE
into IE reversibly, presumably in the form of acoustic waves, and finally
removing the latter by thermal dissipation. The problem with this 
hypothesis, however, is that thermal dissipation is generally found 
to be a significantly less effective way to dissipate acoustic waves 
in liquids than the bulk viscosity, e.g., \cite{Lighthill1978b}
(thermal dissipation is more effective in gases, however).
However, as the bulk viscosity dissipates the divergent velocity 
component, it is absent as a dissipation mechanism in the Boussinesq
and anelastic approximations. If the physical mechanisms underlying
the AGPE dissipation into IE are indeed related to the dissipation
mechanisms of acoustic waves, then it is probably not possible to 
ascertain that only molecular diffusion is involved in reality.

\par

 The overall conclusion is that the diabatic effects due to 
molecular diffusive processes in turbulent stratified fluids seem
to give rise to nontrivial and potentially large conversions between
mechanical energy and internal energy even in fluids traditionally
regarded as incompressible or nearly incompressible, and that such
conversions are actually supported by such models as the Boussinesq
and anelastic approximations even if this is still largely 
unappreciated. The main consequences is that thermodynamics, 
compressible effects, and the divergent component of the fluid
velocity play a potentially more important role than traditionally
assumed for understanding the physical processes and mechanisms 
ultimately involved in the energetics of turbulent mixing in stratified
fluids. The present results, which suggest that the Boussinesq and
anelastic approximations can support large conversions between 
mechanical energy, help rationalise why such approximations appears
to do so well in simulating turbulent stratified flows. On the other
hand, the present results also suggest that real turbulent stratified
fluids should exhibit differences with Boussinesq and anelastic fluids,
as there must be a limit beyond which neglecting the effects of a 
divergent velocity may become noticeable for instance.
 Making progress toward 
clarifying these issues will probably require further detailed analysis
of the energetics of the compressible Navier-Stokes equations along
the lines recently developed by \cite{Tailleux2009}, and direct numerical
simulations of turbulent mixing in fully compressible stratified 
liquid flows resolving acoustic waves emitted by molecular 
diffusive heating/cooling and their dissipation mechanisms. 
Laboratory experiments, of the kind illustrated in Fig. \ref{set_up},
might also help.

\appendix

\bibliographystyle{jfm}

\begin{thebibliography}{13}
  \bibitem[Alberty (2001)]{Alberty2001}
    \textsc{Alberty, R. A.} 2001
    Use of Legendre transforms in chemical thermodynamics.
    \emph{Pure Appl. Chem.} \textbf{73}, 1349--1380.
  \bibitem[Andrews (1981)]{Andrews1981}
    \textsc{Andrews, D. G.} 1981
    A note on potential energy density in a stratified compressible fluid.
    \emph{J. Fluid Mech.} \textbf{107}, 227--236.
  \bibitem[Bannon (1995)]{Bannon1995}
  \textsc{Bannon, P. R.} 1995
    Hydrostatic adjustment: Lamb's problem.
    \emph{J. Atm. Sciences} \textbf{52}, 2302--2312.
   \bibitem[Bannon (1996)]{Bannon1996}
      \textsc{Bannon, P.R.} 1996
      On the anelastic approximation for a compressible atmosphere.
      \emph{J. Atm. Sciences} \textbf{53}, 3618--3628.
  \bibitem[Boussinesq (1903)]{Boussinesq1903}
      \textsc{Boussinesq, J.} 1903
      Th\'{e}orie analytique de la chaleur. Vol 2. Gauthier-Villars, Paris.
   \bibitem[Bryan (1969)]{Bryan1969}
      \textsc{Bryan, K.} 1969
      A numerical model for the study of the circulation of the world ocean.
      \emph{J. Comp. Phys.} \textbf{7}, 347--376.
    \bibitem[Callen (1985)]{Callen1985}
       \textsc{Callen, H.B.} 1985
       Thermodynamics and an introduction to thermostatistics. 
       Wiley. 493 pp. 
    \bibitem[Davies \& et al. (2003)]{Davies2003}
        \textsc{Davies, T., Staniforth, A., Wood, N. \&
        Thuburn, J.} 2003
        Validity of anelastic and other equation sets as inferred from
        normal-mode analysis. \emph{Q. J. Roy. Meteorol. Soc.} 
        \textbf{129}, 2761--2775.
   \bibitem[de Groot \& Mazur (1962)]{degroot1962}
     \textsc{de Groot, S.R. \& Mazur, P.} 1962
      Non-equilibrium thermodynamics. North Holland Publishers.
   \bibitem[Durran (1989)]{Durran1989}
     \textsc{Durran, D. R.} 1989
     Improving the anelastic approximation. 
     \emph{J. Atmos. Sci.} \textbf{46}, 1453--1461.
   \bibitem[de Szoeke \& Samelson (2002)]{Deszoeke2002}
     \textsc{de Szoeke, R. A. \& Samelson, R. M.} 2002
     The duality between Boussinesq and non-Boussinesq hydrostatic
    equations of motion. \emph{J. Phys. Oceanogr.} \textbf{30},
     2194--2203.
   \bibitem[Feistel (2003)]{Feistel2003}
      \textsc{Feistel, R.} 2003
      A new extended Gibbs thermodynamic potential of seawater.
      \emph{Prog. Oceanogr.} \textbf{58}, 43--114.
   \bibitem[Griffies (2004)]{Griffies2004}
       \textsc{Griffies, S. M.} 2004
       Fundamentals of Ocean Climate Models. 
       Princeton University Press.
   \bibitem[Holliday \& McIntyre (1981)]{Holliday1981}
       \textsc{Holliday, D. \& McIntyre, M. E.} 1981
        On potential energy density in an incompressible stratified fluid.
       \emph{J. Fluid Mech.} \textbf{107}, 221--225.
  \bibitem[Hughes \& Griffiths (2008)]{Hughes2008}
        \textsc{Hughes, G. 0. \& Griffiths, R. W.} 2008
        Horizontal convection.
       \emph{Annu. Rev. Fluid Mech.} \textbf{40}, 185--208.
   \bibitem[Hughes \& al. (2009)]{Hughes2009}
        \textsc{Hughes, G.0., Hogg, A., \& Griffiths, R.W.} 2009
        Available potential energy and irreversible mixing in the
        meridional overturning circulation. 
        \emph{J. Phys. Oceanogr.} \textbf{39}, 3130--3146.
   \bibitem[Ingersoll \& Pollard (1982)]{Ingersoll1982}
       \textsc{Ingersoll, A. P. \& Pollard, D.} 1982
        Motions in the interiors and atmospheres of Jupiter and 
        Saturn: Scale analysis, anelastic equations, barotropic
        stability criterion. \emph{Icarus} \textbf{52}, 62--80.
   \bibitem[Ingersoll (2005)]{Ingersoll2005}
        \textsc{Ingersoll, A.} 2005
       Boussinesq and anelastic approximations revisited:
        Potential energy release during thermobaric instability.
        \emph{J. Phys. Oceanogr.} \textbf{35}, 1359--1369.
      \bibitem[Lighthill (1978a)]{Lighthill1978a}
          \textsc{Lighhill, M. J.} 1978a
          `Acoustic stream', \emph{J. Sound Vibr.} \textbf{61}, 391--418.
      \bibitem[Lighthill (1978b)]{Lighthill1978b}
           \textsc{Lighthill, M.J.} 1978b
            Waves in fluids, Cambridge University Press.   
    \bibitem[Lipps \& Hemler (1982)]{Lipps1982}
         \textsc{Lipps, F. B. \& Hemler, R. S.} 1982
         A scale analysis of deep moist convection and some related
         numerical calculations. \emph{J. Atmos. Sci.} \textbf{39}, 2192--2210.
    \bibitem[Klein (2009)]{Klein2009}
        \textsc{Klein, R.} 2009
        Asymptotics, structure, and integration of sound-proof 
        atmospheric flow equations.
        \emph{Theor. \& Comput. Fluid Dyn.} \textbf{23}, 161--195.
    \bibitem[Klein (2010)]{Klein2010}
        \textsc{Klein, R.} 2010
        Scale-dependent models for atmospheric flows.
        \emph{Annu. Rev. Fluid Mech.} \textbf{42}, 249--274.
    \bibitem[Lilly (1996)]{Lilly1996}
          \textsc{Lilly, D.K.} 1996
          A comparison of incompressible, anelastic and Boussinesq
         dynamics. \emph{Atmos. Res.}, \textbf{40}, 143--151.
%
   \bibitem[Lorenz (1955)]{Lorenz1955}
        \textsc{Lorenz, E. N.} 1955
        Available potential energy and the maintenance of the general
        circulation. \emph{Tellus} \textbf{7}, 157--167.
   \bibitem[McIntyre (2010)]{McIntyre2010}
       \textsc{McIntyre, M. E.} 2010
        On spontaneous imbalance and ocean turbulence: generalizations
        of the Paparella-young epsilon theorem. In
        Turbulence in the Atmosphere and Oceans. Proc.
        International IUTAM/Newton Inst. Workshop held 
        8-12 December 2008, ed. D. G. Dritschel, Springer-Verlag.
   \bibitem[McDougall (2003)]{McDougall2003}
        \textsc{McDougall, T.J.} 2003
        Potential enthalpy: A conservative oceanic variable for evaluating
        heat content and heat fluxes.
        \emph{J. Phys. Oceanogr.} \textbf{33}, 945--963. 
   \bibitem[M\"{u}ller (1998)]{Muller1998}
       \textsc{M\"{u}ller, B.} 1998
        Low-Mach number aymptotics of the Navier-Stokes equations
        \emph{J. Engineering Math.} \textbf{34}, 97--109.
    \bibitem[Nycander \& al. (2007)]{Nycander2007}
       \textsc{Nycander, J., Nilsson, J., 
      D\"{o}\"{o}s, \& Brostr\"{o}m, G.} 2007
       Thermodynamic analysis of the ocean circulation.
       \emph{J. Phys. Oceanogr.} \textbf{37}, 2038--2052.
     \bibitem[Nycander (2010)]{Nycander2010}
        \textsc{Nycander, J.}{2010}
        Horizontal convection with a nonlinear equation of state:
        generalization of a theorem of Paparella and Young.
        \emph{Tellus} \textbf{62A}, 134--137. 
        \bibitem[Ogura \& Phillips (1962)]{Ogura1962}
       \textsc{Ogura, Y. \& Phillips, N. A.} 1962
       Scale analysis of deep and shallow convection in the atmosphere.
       \emph{J. Atmos. Sci.} \textbf{19}, 173--179.
     \bibitem[Oberbeck (1879)]{Oberbeck1879}
       \textsc{Oberbeck, A.} 1879
         \"{U}ber die W\"{a}rmeleitung der Fl\"{u}ssigkeiten bei
         Ber\"{u}cksichtigung der Str\"{o}mungen infolge vor
         Temperaturedifferenzen (On the thermal conduction of liquids
         taking account flows due to temperature differences).
         \emph{Ann. Phys. Chem., Neue Folge.} \textbf{7}, 271--292.
   \bibitem[Paparella \& Young (2002)]{Paparella2002}
       \textsc{Paparella, F. \& Young, W. R.} 2002
       Horizontal convection is non turbulent.
       \emph{J. Fluid Mech.} \textbf{466}, 205-214.
   \bibitem[Pauluis (2008)]{Pauluis2008}
       \textsc{Pauluis, 0.} 2008
       Thermodynamic consistency of the analestic approximation for
       a moist atmosphere. \emph{J. Atm. Sc.}, \textbf{65}, 
       2719--2729.
  \bibitem[Pons \& Le Qu\'er\'e (2005)]{Pons2005}
        \textsc{Pons, M. \& P. Le Qu\'er\'e} 2005
        An example of entropy balance in natural convection.
        Part 2: the thermodynamic Boussinesq equations.
       \emph{C. R. Mecanique} \textbf{333}, 133--138.
  \bibitem[Pons \& Le Qu\'er\'e (2007)]{Pons2007}
        \textsc{Pons, M. \& P. Le Qu\'er\'e} 2007
        Modeling natural convection with the work of pressure-forces,
      a thermodynamic necessity. 
      \emph{Int. J. Numer. Meth. Heat Fluid Flow} \textbf{17}, 322--332. 

    \bibitem[Shchepetkin \& McWilliams (2011)]{Shchepetkin2011}
        \textsc{Shchepetkin A. F. \& McWilliams, J. C.} 2011
        Accurate Boussinesq oceanic modelling with a practical,
        ``stiffened'' equation of state.
       \emph{Ocean Modelling}, \textbf{38}, 41--70.
   \bibitem[Spiegel \& Veronis (1960)]{Spiegel1960}
       \textsc{Spiegel, E. A. \& Veronis, G.} 1960
       On the Boussinesq approximation for a compressible fluid.
       \emph{Astrophys. Journal} \textbf{131}, 442--447.
    \bibitem[Tailleux (2009)]{Tailleux2009}
        \textsc{Tailleux, R.} 2009
        On the energetics of stratified turbulent mixing, irreversible
        thermodynamics, Boussinesq models, and the ocean heat engine
        controversy. \emph{J. Fluid Mech.} \textbf{638}, 339--382.
     \bibitem[Tailleux \& Rouleau (2010)]{Tailleux2010c}
        \textsc{Tailleux, R. \& Rouleau, L.} 2010
        The effect of mechanical stirring on horizontal convection.
        \emph{Tellus A} \textbf{62}, 138--153.
     \bibitem[Tailleux (2010)]{Tailleux2010}
        \textsc{Tailleux, R.} 2010
        Identifying and quantifying nonconservative energy 
        production/destruction terms in hydrostatic Boussinesq 
        primitive equation models. \emph{Ocean Modell.}, \textbf{34},
        125--136.
     \bibitem[Tailleux (2010b)]{Tailleux2010b}
        \textsc{Tailleux, R.} 2010
        On the buoyancy power input in the oceans energy cycle.
        \emph{Geophys. Res. Lett.}, \textbf{37}, 
         L22603, doi:10.1029/2010GL044962.
     \bibitem[IOC (2010) ]{teos10}
        \textsc{IOC, SCOR and IAPSO} 2010
        The international thermodynamic equation of seawater - 2010:
      Calculation and use of thermodynamic properties.
      Intergovernmental Oceanogrphic Commission, Manuals and Guides
      No. 56, UNESCO (English), 196 pp. 
   \bibitem[Vallis (2006)]{Vallis2006}
      \textsc{Vallis, G.}{2006}
      Atmospheric and oceanic fluid dynamics.
      Cambridge University Press.
      \bibitem[Vanneste \& B\"{u}hler (2011)]{Vanneste2001}
      \textsc{Vanneste, J. \& O. B\"{u}hler} 2011
      Streaming by leaky surface acoustic waves.
      \emph{Proc. R. Soc. Lond. A}, \textbf{467}, 1779--1800.
   \bibitem[Wang \& Huang (2005)]{Wang2005}
      \textsc{Wang \& Huang, R. X.}{2005}
      An experimental study on thermal circulation driven by horizontal
      differential heating.
     \emph{J. Fluid Mech.} \textbf{540}, 49--73.
   \bibitem[Winters \& al (1995)]{Winters1995}
       \textsc{Winters, K. B., Lombard, P. N., and Riley, J. J., 
       \& d'Asaro (1995)}
       Available potential energy and mixing in density-stratified fluids.
       \emph{J. Fluid Mech.} \textbf{289}, 115--228.
   \bibitem[Winters \& Young (2009)]{Winters2009}
       \textsc{Winters, K.B. \& Young, W.R.}{2009}
       Available potential energy and buoyancy variance in horizontal
       convection. \emph{J. Fluid Mech.} \textbf{629}, 221--230.
     \bibitem[Young (2010)]{Young2010}
        \textsc{Young, W. R.}{2010}
        Dynamic enthalpy, conservative temperature, and the seawater
        Boussinesq approximation. \emph{J. Phys. Oceanogr.}, 
        \textbf{40}, 394--400.
\end{thebibliography}

\end{document}